\documentclass[draftclsnofoot, onecolumn, 11pt]{IEEEtran}
\usepackage{fancyhdr}
\usepackage{amsfonts}
\usepackage{times}
\usepackage{latexsym}
\usepackage{dsfont}
\usepackage{cite}
\usepackage{verbatim}
\usepackage{subfigure}
\newtheorem{theorem}{Theorem}

\newtheorem{assumption}[theorem]{Assumption}

\newtheorem{definition}[theorem]{Definition}

\newtheorem{proposition}[theorem]{Proposition}

\newcommand{\bydef}{\triangleq}

\def\bydef{:=}

\def\bb0{{\mathbb{0}}}

\def\bydef{:=}
\def\ba{{\mathbf{a}}}
\def\bb{{\mathbf{b}}}

\def\bee{{\mathbf{e}}}
\def\bff{{\mathbf{f}}}
\def\bg{{\mathbf{g}}}
\def\bh{{\mathbf{h}}}

\def\bp{{\mathbf{p}}}

\def\bu{{\mathbf{u}}}
\def\bv{{\mathbf{v}}}
\def\bw{{\mathbf{w}}}
\def\bx{{\mathbf{x}}}
\def\by{{\mathbf{y}}}
\def\bz{{\mathbf{z}}}
\def\b0{{\mathbf{0}}}
\def\bbeta{{\boldsymbol\beta}}

\def\bA{{\mathbf{A}}}
\def\bB{{\mathbf{B}}}

\def\bF{{\mathbf{F}}}

\def\bH{{\mathbf{H}}}
\def\bI{{\mathbf{I}}}

\def\bR{{\mathbf{R}}}

\def\bU{{\mathbf{U}}}

\def\bydef{:=}

\def\sf0{{\mathsf{0}}}

\newcommand{\defeq}{\bydef}

\usepackage{cite}
\usepackage{graphicx}
\usepackage{psfrag}
\usepackage{subfigure}
\usepackage{amsmath}
\usepackage{amssymb}
\usepackage{epsfig}
\usepackage{color}
\usepackage{epsf}
\usepackage{algorithmic}
\usepackage{algorithm}
\usepackage{epsfig}
\usepackage{fancyhdr}
\usepackage{setspace}
\usepackage{wasysym}
\usepackage{sansmath}

\DeclareMathOperator*{\argmin}{arg\,min}
\DeclareMathOperator*{\argmax}{arg\,max} 

\def\Nsc{{N_\mathrm{sc}}}         %
\def\ep{{{\bee_\mathrm{p}[t]}}}
\def\Rsum{\bar{R}_\mathrm{sum}}
\def\fd{{f_\mathrm{D}}}         %
\def\Ts{{T_\mathrm{s}}}         %
\def\sigman{{\sigma_\mathrm{n}^2}}         %
\def\Cmag{{\mathcal{C}_\mathrm{mag}}}         %
\def\Cdir{{\mathcal{C}_\mathrm{dir}}}         %
\def\Ndir{{N_\mathrm{dir} }}         %
\def\Nmag{{N_\mathrm{mag}}}         %
\def\emax{{\widehat{e}_\mathrm{max}}}  
\def\emin{{\widehat{e}_\mathrm{min}}}  
\def\eavg{{\widehat{e}_\mathrm{avg}}} 
\def\emag{{\widehat{e}_\mathrm{mag}}} 
\def\edir{{\widehat{\bee}_\mathrm{dir}[t]}} 
\def\etaf{{\eta_\mathrm{f}}} 

\begin{document}

\title{Grassmannian Differential Limited Feedback for Interference Alignment\thanks{This work was supported by the Office of Naval Research (ONR) under grant N000141010337.}}

\author{\IEEEauthorblockN{Omar El Ayach and Robert W. Heath, Jr.\\}
\IEEEauthorblockA{The University of Texas at Austin\\
Department of Electrical \& Computer Engineering\\
1 University Station C0806\\
Austin, TX, 78712-0240 \\
Email: \{omarayach, rheath\}@mail.utexas.edu}}

\maketitle


\begin{abstract}

Channel state information (CSI) in the interference channel can be used to precode, align, and reduce the dimension of interference at the receivers, to achieve the channel's maximum multiplexing gain, through what is known as interference alignment. Most interference alignment algorithms require knowledge of all the interfering channels to compute the alignment precoders. CSI, considered available at the receivers, can be shared with the transmitters via limited feedback. When alignment is done by coding over frequency extensions in a single antenna system, the required CSI lies on the Grassmannian manifold and its structure can be exploited in feedback. Unfortunately, the number of channels to be shared grows with the square of the number of users, creating too much overhead with conventional feedback methods. This paper proposes Grassmannian differential feedback to reduce feedback overhead by exploiting both the channel's temporal correlation and Grassmannian structure. The performance of the proposed algorithm is characterized both analytically and numerically as a function of channel length, mobility, and the number of feedback bits. The main conclusions are that the proposed feedback strategy allows interference alignment to perform well over a wide range of Doppler spreads, and to approach perfect CSI performance in slowly varying channels. Numerical results highlight the trade-off between the frequency of feedback and the accuracy of individual feedback updates.
\end{abstract}

\newpage


\section{Introduction} \label{sec:intro}

Interference alignment (IA) is a precoding strategy that attempts to align interfering signals in time, frequency, or space. By aligning signals, IA reduces the dimension of the interference observed at each receiver and achieves the channel's maximum multiplexing gain. With alignment, network sum rate can increase linearly with the number of communicating users~\cite{Cadambe2008}. Various methods of calculating IA precoders have been proposed~\cite{Peters2010, Gomadam2008, MMSE-IA, Berry-BidirectionalIA, dimakis, Choi2009, santamaria-max-sum-rate, Tresch}. While ~\cite{Peters2010,Gomadam2008, MMSE-IA, Berry-BidirectionalIA, dimakis, Choi2009, santamaria-max-sum-rate, Tresch} are algorithmically different, they share the common feature that channel state information (CSI) for the interference channels is required to compute the IA solution. In network architectures envisioned for IA, the channels would be known at a centralized processor that computes the IA precoders and distributes the solution to the transmitters, or each transmitter would have joint knowledge of all the CSI to compute the solution independently. In either case, a reasonable approach for satisfying the CSI requirement is to feedback quantized CSI through limited feedback, of which a detailed overview can be found in \cite{love2008overview} and the references therein. A straightforward application of limited feedback, however, incurs significant overhead in IA networks. The reasons are that (i) feedback about every interfering link is required from every receiver and (ii) high resolution CSI is required to achieve alignment. As a result, low-overhead methods to transfer CSI must be designed to realize the data rates promised by IA.


Several approaches have been proposed to reduce feedback overhead in IA networks~\cite{Peters2010a, guillaud2011interference, Thukral2009, Krishnamachari2009, Ayach2010a}. In \cite{Peters2010a}, knowledge of channel gains is used to partition the network into optimally sized alignment sub-groups. In~\cite{guillaud2011interference}, IA is applied to partially connected interference channels. Exploiting the network structure as in \cite{Peters2010a, guillaud2011interference} limits the number of channels that must be fed back but does not provide better compression or higher quality CSI. To reduce the number of bits needed to feedback a channel response, \cite{Thukral2009} uses Grassmannian codebooks to quantize the channel coefficients in single-input single-output (SISO) systems with frequency extensions. The work in~\cite{Krishnamachari2009} extends that of \cite{Thukral2009} to systems with multiple antennas. Both~\cite{Thukral2009} and~\cite{Krishnamachari2009} show that IA with limited feedback can achieve a system's degrees of freedom as long as feedback bits scale sufficiently with SNR. As the number of feedback bits increases, however, complexity increases and limited feedback becomes less practical as very large codebooks are undesirable. Analog feedback overcomes the problem of scaling complexity since the CSI distortion decreases naturally with SNR~\cite{Ayach2010a}. Analog feedback, however, is only optimal when the number of feedback channel uses matches the number of symbols being fed back~\cite{caire710multiuser}. Further, a main limitation of the feedback strategies in \cite{Thukral2009,Krishnamachari2009, Ayach2010a} is that they neglect the temporal correlation in the channel that can be exploited to further reduce overhead and improve CSI quality.

Differential and predictive feedback strategies have been proposed for compressing various forms of CSI in time-varying channels~\cite{lee2009mimo, trivellato2008transceiver, sacristn2010differential, sacristn2009differential, love-utilizing-temporal-correlation, RohRao_efficientfeedback, jafarkhani, Takao2}. Simplified differential strategies, such as \cite{lee2009mimo}, have been considered in commercial wireless standards to feedback channel quality indicators. In \cite{sacristn2010differential} differential feedback of multiple-input multiple-output (MIMO) channel correlation matrices is proposed by leveraging the geometry of positive definite Hermitian matrices. The strategy in \cite{sacristn2010differential} was shown to work well when implemented on a wireless testbed~\cite{sacristn2009differential}. In \cite{love-utilizing-temporal-correlation} the differential feedback of unitary precoders is considered in a single-user MIMO setting. The geometry of IA's required CSI, however, neither fits the structure considered in \cite{lee2009mimo,sacristn2009differential,sacristn2010differential, love-utilizing-temporal-correlation}, nor the parametrization derived in \cite{RohRao_efficientfeedback}. Thus, the feedback strategies in \cite{sacristn2009differential,sacristn2010differential, lee2009mimo, love-utilizing-temporal-correlation, RohRao_efficientfeedback} cannot be used. In \cite{trivellato2008transceiver}, tree-structured beamforming codebooks are constructed allowing receivers to feedback differentially encoded information about a local neighborhood in the feedback tree. In \cite{jafarkhani} differential feedback is applied to beamforming vectors that are gradually rotated on the Grassmann manifold using a Householder matrix to improve quantization accuracy. In \cite{Takao2} Grassmannian beamforming vectors are fed back and reconstructed using predictive vector quantization tools, leveraging manifold concepts such as tangent vectors and geodesic curves. Unlike the single-user system considered in \cite{jafarkhani}, however, IA systems are sensitive to leakage interference, and require more feedback than the broadcast channel~\cite{trivellato2008transceiver,Takao2}. As a result, the application of \cite{jafarkhani,Takao2, trivellato2008transceiver} to IA networks is non-trivial and it is not clear whether they can achieve sufficient performance with practical codebook sizes. 

In this paper, we propose a limited feedback strategy based on Grassmannian differential feedback to fulfill the CSI requirement in wideband single antenna systems using IA over frequency extensions. Frequency extensions provide a practical source of dimensionality for IA precoding in SISO interference channels and play an integral role in achieving the maximum multiplexing gain in both single and multiple antenna systems~\cite{Cadambe2008}. Although we present results for single antenna systems, the feedback strategy can be generalized to systems with multiple antennas in a manner similar to \cite{Krishnamachari2009}. The proposed strategy tracks the slow evolution of the normalized channel impulse responses on the Grassmannian manifold. At each feedback update, a quantized tangent vector relating consecutive channel realizations is used at the transmitter to reconstruct the channel. Since the tangent space geometry changes at each iteration, and varying Doppler spreads lead to different channel dynamics, we construct quantization codebooks that adapt to the channel's geometry and dynamics. Unlike \cite{Takao2}, no prediction is performed and differential feedback is used in the new context of SISO interference channels. We show that the performance of the proposed differential strategy cannot be improved upon by a simple yet commonly used class of linear or geodesic predictors. Thus, the development of more sophisticated prediction algorithms that can further improve CSI quality remains a promising venue for future work. 

We derive an approximation for the average distortion achieved by the proposed algorithm and show that it is accurate for a wide range of channel and feedback parameters. Unlike the analysis of the Householder strategy in \cite{jafarkhani} and the predictive strategy in \cite{Takao2}, which neglect either the quantization process or the channel's Doppler spread, the derived performance characterization captures the effects of both fading speed and codebook size. Simulation results show that the proposed algorithm outperforms memoryless quantization and other competitive feedback strategies for temporally correlated channels \cite{Takao2}. Through simulation, we highlight also the trade-off between feedback refresh rate and codebook size \cite{love-refreshrate}. We show that, from a CSI distortion perspective, frequent updates using smaller codebooks may often be preferred to less-frequent updates with a large codebook. Finally, while scaling codebook size is still required to preserve multiplexing gain, we show that IA with the proposed feedback strategy performs well in the range of medium to high SNR with a relatively small number of feedback bits.

This paper is organized as follows. Section \ref{sec:sysmodel} introduces the SISO wideband interference channel model with orthogonal frequency division multiplexing (OFDM). Section \ref{sec:ia} reviews the concept of interference alignment over frequency extended channels and highlights the degradation due to imperfect channel knowledge. Section \ref{sec:gdc} presents the proposed feedback framework while Section \ref{sec:gdc_design} elaborates on its main design parameters. Sections \ref{sec:approximation} and \ref{sec:sims} present analytical and numerical results respectively on the performance of the Grassmannian differential feedback, as well as the performance of IA when the proposed strategy is used for CSI feedback. We conclude with Section \ref{sec:conclusion}.

We summarize the notation employed in this paper: $\bA$ is a matrix; $\ba$ is a vector; and $a$ is a scalar; $(\cdot)^*$ denotes the conjugate transpose; $\bA \circ \bB$ is the Hadamard product; $\|\bA\|_F$ is the Frobenius norm of $\bA$ and $\mathrm{trace}(\bA)$ is its trace; $\|\ba\|$ is the $2$-norm of $\ba$; $\left|a\right|$ is the absolute value of $a$; $\bI_N$ is the $N\times N$ identity matrix; $\mathbf{0}_{N}$ is the N-dimensional zero vector; $\boldsymbol{\mathcal{F}}_{N}$ is the $N$-point discrete Fourier transform (DFT) matrix; $\mathrm{diag}(\ba)$ is the diagonal matrix obtained by putting the elements of $\ba$ on its diagonal; $[\ba, \bb]$ is a horizontal concatenation of vectors $\ba$ and $\bb$, $\mathbb{C}^N$ is the $N$-dimensional complex space; $\mathcal{CN}(\ba,\bA)$ is a complex Gaussian random vector with mean $\ba$ and covariance matrix $\bA$. Expectation is denoted by $\mathbb{E}\left[\cdot\right]$. 
\section{System Model} \label{sec:sysmodel}

Consider a frequency selective SISO interference channel with $K$ communicating user pairs. Each user $k$ communicates desired data to its paired receiver $k$ and interferes with all other receivers $\ell \neq k$. The wideband channel between transmitter $\ell$ and receiver $k$ at time $t$ is given by the $L$-tap channel impulse response vector $\bh_{k,\ell}[t]=\left[h_{k,\ell}[t,0]^*,h_{k,\ell}[t,1]^*,\hdots,h_{k,\ell}[t,L-1]^*\right]^*  \ \forall k,\ell \in \{1, \hdots, K\}$. Channel impulse responses are drawn independently across $k$ and $\ell$ from a continuous distribution and have covariance matrix $\mathbb{E}\left[\bh_{k,\ell}[t]\bh_{k,\ell}[t]^*\right] =\mathbf{R}_{\bh_{k,\ell}}, \ \forall k,\ell \in\left\{1, \hdots, K\right\}$ and $\forall t$. Letting $\rho_{k,\ell}$ be the average attenuation of user $\ell$'s signal at user $k$'s receiver, the channel response satisfies $\mathrm{trace}\left(\mathbf{R}_{\bh_{k,\ell}} \right)=\rho_{k,\ell}$. In the case of uncorrelated channel taps, $\mathbf{R}_{\bh_{k,\ell}}=\mathrm{diag}\left(\bp_{k,\ell}\right)$, where $\bp_{k,\ell}$ is the $L\times 1$ vector representing the channel's power delay profile. 


Using orthogonal frequency division multiplexing (OFDM), the transmitters transform the observed frequency selective channels into a set of $\Nsc$ non-interfering narrowband subcarriers. Stacking each received OFDM symbol in a vector, the input-output relationship can be written in matrix form as
\begin{equation}
\by_k[t]=\bH_{k,k}[t]\bx_k[t]+\sum\limits_{\ell \neq k}\bH_{k,\ell}[t]\bx_\ell[t] + \bv_k[t],
\label{eqn:ofdm_sig_model}
\end{equation}
where $\bx_k[t]$ is the OFDM symbol sent by user $k$ at time $t$ with the average power constraint $\mathbb{E}\left[\|\bx_k[t]\|^2\right]=\Nsc P$, the $\Nsc \times \Nsc$ matrix $\bH_{k,\ell}[t]=  \mathrm{diag}\left(\boldsymbol{\mathcal{F}}_{\Nsc}\left[\bh_{k,\ell}^*[t],\ \mathbf{0}_{\Nsc-L}\right]^*\right)$ represents the channel frequency response between transmitter $\ell$ and receiver $k$ at time $t$, and $\bv_k[t]$ is the i.i.d. $\mathcal{CN}(0,\sigman \bI_{\Nsc})$ thermal noise observed by user $k$. The system model assumes perfect time and frequency synchronization, and a cyclic prefix that is long enough to accommodate the impulse response of all channels as well as the potential propagation delay between users.

Further, the relationship in (\ref{eqn:ofdm_sig_model}) implicitly assumes that the $L$-tap channels $\bh_{k,\ell}[t]$ seen by the $t^\textrm{th}$ OFDM symbol remain constant over the OFDM symbol time and pilots can be used to estimate them at the receiver. 
The channels over consecutive OFDM symbols, however, are considered to be slowly varying. Let $\fd $ be the Doppler frequency of the channel and $\Ts$ be the OFDM symbol time. We assume channels are temporally correlated according to the model proposed by Clarke \cite{clarke1968statistical}, i.e $\mathbb{E}\left[\rho_{k,\ell}^{-1}\left|\bh_{k,\ell}^*[t-m]\bh_{k,\ell}[t]\right|\right]= J_0(2\pi \fd  \Ts m)$ where  $J_0$ is the $0$-th order Bessel function of the first kind. A main underlying assumption, however, is that while the channels vary slowly over time, their length $L$ remains fixed. Since channel response length, is likely to vary slower than the channel response itself, and since it is constrained to be a small integer, the channel length can be fed back to the transmitter if necessary at a small extra overhead cost. In general, the proposed feedback algorithm can be applied to all correlated channels, however, further details on the mathematical structure of the temporally correlated channel are given whenever a result is model-dependent. 

Throughout the remainder of this paper we make the following two system-level assumptions:
\begin{itemize}
\item We assume that the channels $\bh_{k,\ell}[t]\ \forall \ell$ are known to receiver $k$. This implicitly assumes that the set of $K$ transmitters coordinate (or orthogonalize) their pilot transmissions allowing receivers to estimate all incoming channels. We make a further simplification by assuming receiver CSI is perfect.
\item We assume each receiver has an error-free logical feedback link to all transmitters. The feedback link from receiver $k$ to transmitter $\ell$ may either be a direct link, or a link through other nodes such as transmitter $k$ and/or a centralized processor. Effectively, this allows obtaining global CSI at at least one node to enable the calculation of IA precoders which, if needed, can be forwarded to the other transmitters. 
\end{itemize}
While the importance of transmitter channel knowledge makes the two assumptions rather standard in the literature on precoding for the interference channel~\cite{Cadambe2008, Choi2009, Krishnamachari2009, Peters2010, Peters2010a, Thukral2009, guillaud2011interference, Ayach2010a, dimakis, santamaria-max-sum-rate}, or interference coordination in general~\cite{gesbert2010multi}, they are by no means trivial. Coordinating training phases, feedback phases, and centralized processing constitutes significant complexity in both cellular and ad-hoc systems. We refer the interested reader to \cite{gesbert2010multi} and the references therein for a detailed discussion of the limitations, the benefits, and the possible solutions that enable coordination in the case of cellular systems.

\section{Interference Alignment in Frequency} \label{sec:ia}

In this section we review the concept of IA over frequency extensions when perfect channel state information is available at the transmitter W then summarize the effect of imperfect transmitter channel knowledge on the performance of IA.

\subsection{IA with Perfect CSI at the Transmitter}  \label{sec:perfect_csi}

IA for the SISO interference channel can achieve the maximum degrees of freedom when coding over infinite channel extensions~\cite{Cadambe2008}.
Using IA over $N$ frequency extensions, each transmitter $k$ at time $t$ sends $d_k < N$ symbols, $x_k^m[t]$, along the $N \times 1$ precoding vectors $\bff_k^m[t]$. As a result, the transmitted symbol is 
\begin{equation}
\bx_k[t]=\sum\limits_{m=1}^{d_k}\bff_k^m[t]x_k^m[t],
\label{eqn:ofdm_prec}
\end{equation}
where $\|\bff_k^m[t]\|_2=1$ and $\mathbb{E}\left[|x_k^m[t]|^2\right]=NP/d_k$, such that the power in each $N$ subcarriers is $NP$~\cite{martinian-waterfilling}. The transmit directions $\bff_k^m[t]$ are calculated such that the interference from $K-1$ users is aligned at all receivers, leaving at least $d_k$ interference free dimensions for the desired signal. The number of transmitted symbols $d_k$ can be chosen according to the original strategy in \cite{Cadambe2008} or the improved method in \cite{Choi2009}, both of which asymptotically achieve the maximum degrees of freedom, i.e. $\lim_{N\longrightarrow\infty}\frac{1}{N}\sum d_k=K/2$.

 Note that users need not code over all $\Nsc$ subcarriers since algorithm complexity increases significantly with the number of subcarriers whereas the achieved multiplexing gain may increase very slowly with $\Nsc$ as demonstrated by the IA solutions in \cite{Cadambe2008}. Users could potentially treat any $N$-subset of subcarriers as an independent coding group and thus have $\lfloor\frac{\Nsc}{N}\rfloor$ parallel alignment groups in an OFDM symbol. In this case (\ref{eqn:ofdm_prec}) applies to each group of subcarriers, and one can write $\lfloor\frac{\Nsc}{N}\rfloor$ relationships for each OFDM symbol. For simplicity of exposition, in the remainder of this section we assume $\lfloor\frac{\Nsc}{N}\rfloor=1$.


To quantify concisely the degradation due to imperfect CSI, we consider a zero-forcing receiver; other receiver designs can be used. At the output of the linear receivers $\bw_k^m[t]$, the received signal is
\begin{align}
\begin{split}
\bw_k^m[t]^*\by_k[t] = &\bw_k^m[t]^*\mathbf{H}_{k,k}[t]\bff_{k}^{m}[t] x_{k}^{m}[t]+\sum\limits_{\ell\neq m}\bw_k^m[t]^*\mathbf{H}_{k,k}[t]\bff_{k}^{\ell}[t] x_{k}^{\ell}[t] \\ 
& + \sum\limits_{i\neq k}\sum\limits_{\ell=1}^{d_i} \bw_k^m[t]^*\mathbf{H}_{k,i}[t]\bff_{i}^{\ell}[t] x_{i}^{\ell}[t]+\bw_k^m[t]^*\bv_k[t],
\label{eqn:recv_proj}
\end{split}
\end{align}
for $m\in\left\{1,\ldots,d_k\right\}$ and $k \in \left\{1,\ldots,K\right\}$, where $\|\bw_k^m[t]\|^2=1$. Assuming a zero-forcing receiver, the conditions for perfect interference alignment are
\begin{eqnarray}
\bw_k^m[t]^*\mathbf{H}_{k,k}[t]\bff_{k}^{\ell}[t]=0 &\quad& \forall k,\  \ell\neq m \label{eqn:conditions1}\\
\bw_k^m[t]^*\mathbf{H}_{k,i}[t]\bff_{i}^{\ell}[t]=0 &\quad& \forall i\neq k,\text{ and } \forall m, \ell \label{eqn:conditions2}\\
\left|\bw_k^m[t]^*\mathbf{H}_{k,k}[t]\bff_{k}^{m}[t]\right|\geq c >0 &\quad& \forall k,m \label{eqn:conditions3}
\label{eqn:conditions}
\end{eqnarray}
where alignment is achieved by (\ref{eqn:conditions1}) and (\ref{eqn:conditions2}), while (\ref{eqn:conditions3}) ensures the decodability of the $d_k$ desired streams.

The sum rate achieved by the linear zero-forcing receiver \cite{makouei-simple-sinr-characterization}, assuming Gaussian input signals, is
\begin{equation}
\Rsum = \sum\limits_{k=1}^{K}\sum\limits_{m=1}^{d_k}
\frac{1}{\Nsc}\log_2\left(1+\frac{\frac{\Nsc P}{d_k}\left|\bw_k^m[t]^*\mathbf{H}_{k,k}[t]\bff_{k}^{m}[t]\right|^2}{\mathcal{I}_{k,m}^1[t]+\mathcal{I}_{k,m}^2[t]+ \sigman }\right),
\label{eqn:sumrate}
\end{equation}
where $\mathcal{I}_{k,m}^1[t]$ is self-interference from other transmit streams and $\mathcal{I}_{k,m}^2[t]$ is the interference from other users. The two interference terms are 
\begin{align}
\begin{split}
\mathcal{I}_{k,m}^1[t] & =  \sum\limits_{\ell\neq m}\frac{\Nsc P}{d_k}\left|\bw_k^m[t]^*\mathbf{H}_{k,k}[t]\bff_{k}^{\ell}[t]\right|^2, \\
\mathcal{I}_{k,m}^2[t] & =  \sum\limits_{i\neq k}\sum\limits_{\ell=1}^{d_i}\frac{\Nsc P}{d_i}\left|\bw_k^m[t]^*\mathbf{H}_{k,i}[t]\bff_{i}^{\ell}[t]\right|^2.
\label{eqn:leakage}
\end{split}
\end{align}
If perfect channel knowledge is available, and the number of symbols $d_k$ are feasible, equations (\ref{eqn:conditions1})-(\ref{eqn:conditions3}) can be satisfied with probability one and thus $\mathcal{I}_{k,m}^1=\mathcal{I}_{k,m}^2=0$. This gives
\begin{align}
\begin{split}
\lim_{P\rightarrow\infty}\frac{\Rsum}{\log_2 P} & = \lim_{P\rightarrow\infty}  \frac{\sum\limits_{k,m}\frac{1}{\Nsc}\log_2\left(1+\frac{\frac{\Nsc P}{d_k}\left|\bw_k^m[t]^*\mathbf{H}_{k,k}[t]\bff_{k}^{m}[t]\right|^2}{\sigman}\right)}{\log_2 P} = \frac{1}{\Nsc}\sum_k d_k \stackrel{\Nsc\rightarrow \infty}{\longrightarrow} \frac{K}{2},
\label{eqn:mux_gain}
\end{split}
\end{align}
which confirms the fact that IA achieves the $K/2$ degrees of freedom of the $K$ user interference channel. 
While the achievability proof in \cite{Cadambe2008} assumed independent fading across subcarriers, \cite{Thukral2009} has claimed that fading on each subcarrier need not be independent provided that the channel impulse response is long enough. 

\subsection{The Effect of Limited Feedback} \label{sec:imperfect_csi}

In practical systems, receivers cannot feedback the channels $\bh_{k,i}[t]$ with infinite precision, thus only a distorted version of  $\bh_{k,i}[t]$ is available when calculating IA precoders. The distortion caused by limited feedback implies that interference will not be perfectly aligned. The power of residual misaligned interference increases along with transmit power and causes the SINR in the desired signal subspace to saturate, which limits sum rate at high SNR~\cite{makouei-simple-sinr-characterization}. As a result, low overhead feedback strategies must be designed to improve CSI accuracy and allow good IA performance. 

To establish the impact of limited feedback on IA sum rate, and to develop insights into the structure of a good quantizer, we summarize the calculations in \cite{Thukral2009, Ayach2010a}. In \cite{Ayach2010a}, it is shown that if the transmitters use the estimated or quantized channels $\widehat{\bH}_{k,i},\ \forall k,i \in \{1,\ 2,\ \hdots,\ k\}$ to calculate IA precoders, then the mean loss in sum rate can be upper bounded by
\begin{equation}
\Delta \Rsum \leq \sum\limits_{k,m}\frac{1}{\Nsc}\log_2\left(1+\frac{\mathbb{E}_{\bH}\left[\mathcal{I}^1_{k,m}+\mathcal{I}^2_{k,m}\right]}{\sigman}\right).
\label{eqn:jensen}
\end{equation}

Using the results from \cite[Section IV-B]{Thukral2009}, the components of the sum leakage interference, $\mathcal{I}^1_{k,m}+\mathcal{I}^2_{k,m}$, can be bounded by 
\begin{align}
\begin{split}
\frac{\Nsc P}{d_i}\left|\widehat{\bw}_k^m[t]^*\mathbf{H}_{k,i}[t]\widehat{\bff}_{i}^{\ell}[t]\right|^2  \leq \frac{\Nsc P}{d_i} \|\widehat{\bw}_k^m[t]\circ\widehat{\bff}_{i}^{\ell}[t]\|^2\|\bh_{k,i}[t]\|^2  \left(1-\left|\frac{\bh_{k,i}[t]^*\widehat{\bh}_{k,i}[t]}{\|\bh_{k,i}[t]\|\|\widehat{\bh}_{k,i}[t]\|}\right|^2\right),
\label{eqn:interference_bound}
\end{split}
\end{align} 
where $\widehat{\bff}_{i}^{\ell}[t]$ and $\widehat{\bw}_k^m[t]$ are the precoders and combiners calculated using imperfect CSI. The bound in (\ref{eqn:interference_bound}) relates the power of leakage interference to the quality of the quantized CSI and suggests mathematical structure that can be exploited by the quantizer. The last term in (\ref{eqn:interference_bound}) indicates that only knowledge of the normalized channels $\frac{\bh_{k,i}[t]}{\|\bh_{k,i}[t]\|}$ is necessary for IA, further the CSI is phase invariant since $\left|\frac{\bh_{k,i}[t]^*\widehat{\bh}_{k,i}[t]}{\|\bh_{k,i}[t]\| \|\widehat{\bh}_{k,i}[t]\|}\right|^2= \left|\frac{e^{j\theta}\bh_{k,i}[t]^* \widehat{\bh}_{k,i}[t]}{\|e^{j\theta}\bh_{k,i}[t]\| \|\widehat{\bh}_{k,i}[t]\|}\right|^2$. This is not surprising since, if an interfering channel is scaled and rotated by $\alpha \in \mathbb{C}$, the interference subspace remains unchanged, i.e., $\mathrm{diag}\left(\boldsymbol{\mathcal{F}}_{\Nsc} \alpha\bh_{k,i}[t]\right)\bff_i^m[t] =\alpha\mathrm{diag} \left(\boldsymbol{\mathcal{F}}_{\Nsc}\bh_{k,i}[t]\right)\bff_i^m[t]$. As a result, the CSI needed for IA with frequency extensions evolves on the manifold of $L$-dimensional unit norm, rotationally invariant vectors otherwise known as the Grassmannian manifold $\mathcal{G}_{L,1}$~\cite{love-heath-limited-feedback-unitary, Mukkavilli, limitedfeedforward,packings-for-MIMO-BC}. Define the chordal distance between two points on the Grassmannian manifold, $\bx_1$ and $\bx_2$, as $d(\bx_1, \bx_2)= \sqrt{1-\left| \bx_1^*\bx_2\right|^2}$. The bound in (\ref{eqn:interference_bound}) can be rewritten as
\begin{equation}
\frac{\Nsc P}{d_i}\left|\widehat{\bw}_k^m[t]^*\mathbf{H}_{k,i}[t]\widehat{\bff}_{i}^{\ell}[t]\right|^2 \leq \frac{\Nsc P}{d_i} \|\widehat{\bw}_k^m[t]\circ\widehat{\bff}_{i}^{\ell}[t]\|^2\ \|\bh_{k,i}[t]\|^2\ d(\bg_{k,i}[t], \widehat{\bg}_{k,i}[t])^2,
\end{equation}
where  $\bg_{k,i}[t] \triangleq \frac{\bh_{k,i}[t]}{\|\bh_{k,i}[t]\|}$ and $\widehat{\bg}_{k,i}[t] \triangleq \frac{\widehat{\bh}_{k,i}[t]}{\|\widehat{\bh}_{k,i}[t]\|}$, showing that leakage interference can be bounded using the chordal distance between the true and quantized channels~\cite{arias1998geometry}.


\section{Grassmannian Differential Feedback} \label{sec:gdc}

In this section, we present the proposed feedback algorithm. Under some conditions, we show that differential quantization, a special case of predictive vector quantization, cannot be improved upon using a simple yet common class of predictors.

\subsection{Differential Feedback Framework}     \label{sec:gdc_algo}

We propose to differentially encode the normalized channel response vectors by exploiting both their Grassmannian structure and temporal correlation. 
The proposed algorithm encodes CSI increments using a tangent vector which defines the geodesic path between consecutive channel realizations.
Our feedback framework uses tools presented in \cite{arias1998geometry}, which were simplified and first used for MISO broadcast channel feedback in \cite{Takao2}. We consider the separate quantization of the slowly varying normalized channel vectors $\bg_{k,i}[t] \triangleq \frac{\bh_{k,i}[t]}{\|\bh_{k,i}[t]\|} \ \forall t \geq 0$; joint quantization may offer higher performance at the expense of additional complexity~\cite{bhagavatula2010limited}. Therefore, we restrict our attention to one of the channels and drop the user subscripts.

\begin{figure}
  \centering
	\includegraphics[width=6.5in]{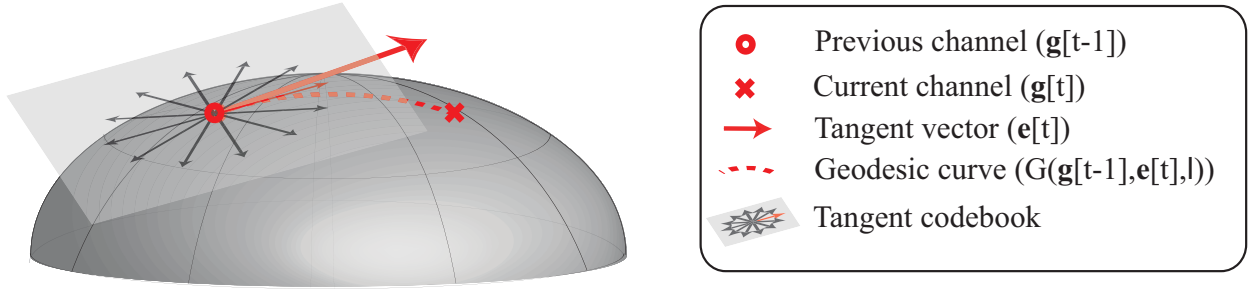}
	\caption{A 3D visualization of the Grassmann manifold and the tangent vector quantization process. The quantized tangent which is selected from the tangent codebook is also marked in red.}
	\label{fig:manifold}
\end{figure}

The smooth structure of the manifold allows each two points to be related using the concept of a geodesic path as shown in Fig. \ref{fig:manifold}. The geodesic curve describing the path between $\bg[t-1]$ and $\bg[t]$ is given by
\begin{equation}
G(\bg[t-1], \bee[t],\ell)=\bg[t-1]\mathrm{cos}\left(\|\bee[t]\|\ell\right) + \frac{\bee[t]}{\|\bee[t]\|}\mathrm{sin}\left(\|\bee[t]\|\ell\right).
\label{eqn:geodesic}
\end{equation}
where the vector $\bee[t]$, also shown in Fig. \ref{fig:manifold}, is known as the tangent vector. The tangent vector can be viewed as a length-preserving unwrapping of the geodesic path onto the tangent space at $\bg[t-1]$. The tangent vector is given by 
\begin{equation}
\bee[t]=\tan^{-1}\left(\frac{d[t]}{\rho[t]}\right)\frac{\bg[t]/\rho[t]-\bg[t-1]}{d[t]/\rho[t]},
\label{eqn:tangent}
\end{equation}
where $\rho[t]=\bg[t-1]^*\bg[t]$ and $d[t]=\sqrt{1-\left|\rho[t]\right|^2}$ \cite{arias1998geometry}. 
From (\ref{eqn:geodesic}), it can be verified that $G(\bg[t-1], \bg[t],0)=\bg[t-1]$, $G(\bg[t-1], \bg[t],1)=\bg[t]$ and that $\|G(\bg[t-1], \bg[t],\ell)\|=1,\ \forall \ell$ since the tangent vector $\bee[t]$ is orthogonal to the base vector $\bg[t-1]$.


\begin{figure}
  \centering
   \includegraphics[width=6.5in]{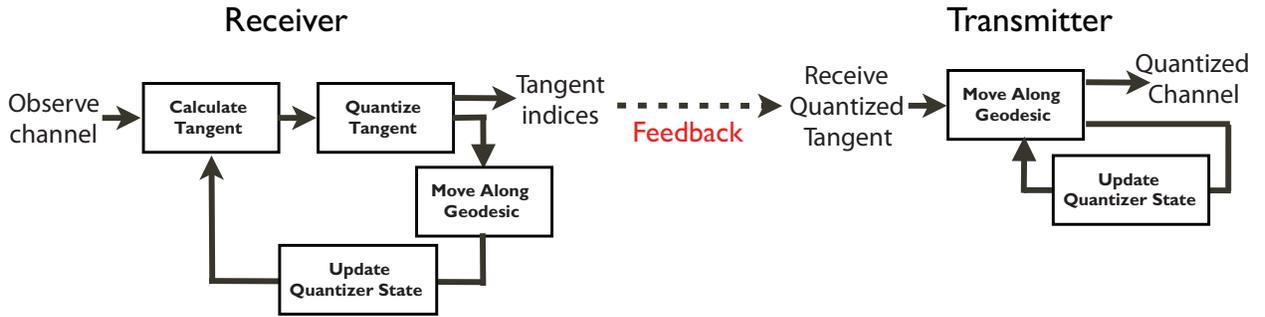}
	\caption{Block diagram of Grassmannian differential quantization and feedback at the transmitter and receiver.}
	\label{fig:block_diag}
\end{figure}

The tangent vector and geodesic path, which together relate two points on the manifold, can be used to build a differential feedback framework to track the evolution of CSI~\cite{Takao2}. Given the previous channel realization and a tangent vector received by feedback, the transmitter can reconstruct the current channel by applying (\ref{eqn:geodesic}). 
The operation of the Grassmannian differential feedback algorithm at both the transmitter and receiver is given in the block diagram of Fig. \ref{fig:block_diag}. For clarity, the pseudo code used to encode the channel evolution over time is given in Algorithm \ref{alg:receiver}. At each new channel realization, the receiver estimates the normalized $L$-tap channel $\bg[t]$. Perfect channel estimation is assumed here to decouple the quantization error from the estimation error. Given the current observation, and the quantized channel in the previous iteration, $\widehat{\bg}[t-1]$, the receiver calculates the tangent vector, $\bee[t]$, from $\widehat{\bg}[t-1]$ to $\bg[t]$. The receiver then quantizes the tangent vector to produce a quantized tangent, $\widehat{\bee}[t]$. The details of the quantization method are given in depth in Section \ref{sec:gdc_design}. This quantized tangent is then fedback to the transmitter over a delay and error free feedback link. Equipped with the previous quantized channel, $\widehat{\bg}[t-1]$, and the quantized error vector $\widehat{\bee}[t]$, \emph{both transmitter and receiver} can now calculate the new quantized channel $\widehat{\bg}[t]$. This vector $\widehat{\bg}[t]$ will later serve as the base point in the next iteration. The algorithm runs in parallel for all channels $\bg_{k,i}[t],\ \forall k,i$ and the transmitters use the estimated CSI vectors $\widehat{\bg}_{k,i}[t],\ \forall k,\ell$ to do interference alignment.

\begin{algorithm}[t!]
\caption{Receiver}
\begin{algorithmic}[1]
\STATE \textbf{Input:} $\widehat{\bg}[t-1]$
\FORALL{$t = 1, 2, \hdots $}
\STATE Estimate the channel $\bg[t]$
\STATE Calculate tangent between $\widehat{\bg}[t-1]$ and $\bg[t]$ using (\ref{eqn:tangent})
\STATE Quantize and feedback the quantized tangent vector $\widehat{\bee}[t]$
\STATE Reconstruct the quantized channel $\widehat{\bg}[t]=G(\widehat{\bg}[t-1], \widehat{\bee}[t],1)$  
\ENDFOR
\end{algorithmic}
\label{alg:receiver}
\end{algorithm}

\begin{algorithm}[t!]
\caption{Transmitter}
\begin{algorithmic}[1]
\STATE \textbf{Input:} $\widehat{\bg}[t-1]$
\FORALL{$t = 1, 2, \hdots $}
\STATE Receive the feedback tangent vector $\widehat{\bee}[t]$
\STATE Reconstruct the quantized channel $\widehat{\bg}[t]=G(\widehat{\bg}[t-1], \widehat{\bee}[t],1)$ 
\ENDFOR
\STATE \textbf{Output:} $\widehat{\bg}[t]$
\end{algorithmic}
\label{alg:transmitter}
\end{algorithm}

\subsection{Comparing Differential vs. Predictive Quantization}    \label{sec:no_pred}

The feedback framework of Section \ref{sec:gdc_algo} is a special case of predictive vector quantization, the main difference being that the proposed strategy does not attempt to further improve CSI by \emph{predicting future channel realizations}. In this section, we begin by formalizing the main difference between predictive and differential feedback. We then show that, for a class of channels, commonly used Grassmannian predictors such as those in~\cite{jafarkhani,Takao2}, yield no performance enhancement over the differential strategy presented. We note, however, that more sophisticated prediction functions may in fact further improve performance. Thus, the development of more intelligent Grasmannian predictors is a promising area for future work.

\subsubsection{Predictive Quantization}
The main idea behind differential feedback is to send the transmitter information that allows it to move from the \emph{old quantized channel} to the \emph{new channel realization}. Thus, a tangent vector $\bee[t]$ is calculated between the base point $\widehat{\bg}[t-1]$ and $\bg[t]$, and is then quantized to give $\widehat{\bee}[t]$. Instead of starting the tangent vector at the point $\widehat{\bg}[t-1]$, general predictive quantization calculates the tangent starting at \emph{another} point $\widetilde{\bg}[t]$, known as the \emph{predicted channel vector}. The premise is that if $\widetilde{\bg}[t]$ is ``closer'' to $\bg[t]$ than $\widehat{\bg}[t-1]$ (the base point used in differential feedback), quantization accuracy can be further improved. The main difficulty now lies in constructing prediction functions that actually yield closer or better estimates of $\bg[t]$.

Unfortunately, work on prediction on the Grassmann manifold is rather limited. In this paper, we consider a class of linear predictors used in \cite{jafarkhani,Takao2}. Though the predictors in \cite{jafarkhani,Takao2} are presented differently, the predicted channel vectors can be written in a unified manner as 
\begin{equation}
\widetilde{\bg}[t]=G(\widehat{\bg}[t-1],\ep ,\ell_\mathrm{p})=\widehat{\bg}[t-1]\cos\left(\|\ep \|\ell_\mathrm{p}\right)+\frac{\ep }{\|\ep \|}\sin\left(\|\ep \|\ell_\mathrm{p}\right), \label{eqn:predictor}
\end{equation}
where $\ep $ is the \emph{predicted tangent vector} and $\ell_\mathrm{p}$ is a predicted step size. Since the norm of $\ep $ can absorb any prediction step size $\ell_P$, we enforce $\ell_\mathrm{p}=1$ without loss of generality.

The challenge now is to use commonly known information, such as $\widehat{\bg}[t^{'}]\  \forall 0\leq t^{'} \leq t-1$,\footnote{In fact \cite{jafarkhani,Takao2} make the simplifying assumption that $\widehat{\bg}[t^{'}]\  \forall 0\leq t^{'} \leq t-1$ is available during predictor design. We relax this assumption, by only allowing access to $\widehat{\bg}[t^{'}]\  \forall 0\leq t^{'} \leq t-1$, at the cost of adding a milder assumption on the quantization error in $\widehat{\bg}[t]$. Results hold, and derivations are simplified, if we revert to the assumptions of \cite{jafarkhani,Takao2}.}  to find the best predicted tangent vector $\bee^\mathrm{opt}_\mathrm{p}[t]$, i.e., the one that maximizes $\mathbb{E}\left[\left|\widetilde{\bg}[t]^*\bg[t]\right|^2\right]$ and makes $\widetilde{\bg}[t]$ the \emph{best estimate} of $\bg[t]$. Thus, the prediction problem can be summarized as solving 
\begin{align}
\bee^\mathrm{opt}_\mathrm{p}[t] & =\argmax_{\ep }\ \mathbb{E}\left[\left|\widetilde{\bg}[t]^*\bg[t]\right|^2\right] \\
& \stackrel{(a)}{=} \argmax_{\ep }\ \mathbb{E}\left[\left|\left(\cos\left(\|\ep \|\right) \widehat{\bg}[t-1]+\sin\left(\|\ep \|\right)\frac{\ep }{\|\ep \|}\right)^*\bh[t]\right|^2\right], \nonumber
\end{align}
where $(a)$ is by replacing $\widetilde{\bg}[t]$ by its definition in (\ref{eqn:predictor}) and by noticing that maximizing $\mathbb{E}\left[\left|\widetilde{\bg}[t]^*\bg[t]\right|^2\right]$ is equivalent to maximizing $\mathbb{E}\left[|\widetilde\bg[t]^*\bh[t]|^2\right]$. In what follows, we show that for a class of temporally correlated channels,  such a geodesic predictor yields no performance enhancement over the differential feedback strategy presented. Therefore, to ensure that prediction does in fact improve performance, more sophisticated predictors that better exploit the channel's Doppler spectrum must be derived. The development of such predictors is a promising topic for future work.

\subsubsection{Performance Comparison \& Prediction Gains}
For the analysis in this section, we make the following assumptions on the channel's temporal structure.
\begin{assumption} 
We assume that the channels $\bh[t]$ are temporally correlated according to a first order autoregressive model~ \cite{baddour2005autoregressive}, and are thus given by $\bh[t]=\etaf\bh[t-1]+\sqrt{1-\etaf^2} \bz[t]$. For simplicity, we further assume that $\bh[t]\sim\mathcal{CN}(0,\frac{1}{L}\bI_L)$ which in turn implies that $\bz[t]\sim\mathcal{CN}(0,\frac{1}{L}\bI_L)$.
\label{as:ar1}
\end{assumption}

While we present our analysis for channels satisfying Assumption \ref{as:ar1}, similar results can be shown for other temporally correlated channel models in which changes in the channel are zero-mean and independent of the previous channel realizations.
\begin{assumption}
We assume that the quantization error introduced by our algorithm is isotropically distributed, i.e., the direction of $\bg[t]-\widehat{\bg}[t]$ is uniformly distributed, and that the error direction and magnitude are independent.
\label{as:quant_err}
\end{assumption}

For a channel following a first order autoregressive model, the predictor's objective function, $\mathbb{E}\left[|\widetilde\bg[t]^*\bh[t]|^2\right]$, can be written as
\begin{align}
\mathbb{E}\left[|\widetilde\bg[t]^*\bh[t]|^2\right] & = \mathbb{E}\left[\left|\left(\cos\left(\|\ep \|\right) \widehat{\bg}[t-1]+\sin\left(\|\ep \|\right)\frac{\ep }{\|\ep \|}\right)^*\left(\etaf \bh[t-1]+\sqrt{1-\etaf^2}\bz[t]\right)\right|^2\right] \nonumber\\
& = \mathbb{E}\left[\left|\etaf\left(\left[\widehat{\bg}[t-1],\ \ep \right]\bbeta\right)^*\bh[t-1]+\bar{\eta}_\mathrm{f} \bbeta^*\bx[t]\right|^2\right],
\end{align}
where we define $\bbeta=[\cos\left(\|\ep \|\right),\ \sin\left(\|\ep \|\right)]^*$, $\bar{\eta}_\mathrm{f}=\sqrt{1-\etaf^2}$ and $\bx[t]=\left[\vphantom{\frac{1}{1}}\widehat{\bg}[t-1]^*,\ \ep^* \right]^*\bz[t]$ to simplify notation. Since $\bz[t]~\sim\mathcal{CN}(0,\frac{1}{L}\bI_L)$ by Assumption \ref{as:ar1}, and since the matrix $\left[\vphantom{\frac{1}{1}}\widehat{\bg}[t-1],\ \ep \right]$ is unitary and independent of $\bz[t]$ by definition, the vector $\bx[t]$ has distribution $\mathcal{CN}(0,\frac{1}{L}\bI_2)$ by the unitary invariance of the Gaussian distribution~\cite{tulino2004random}. Expanding the predictor's objective function, we get
\begin{align}
\begin{split}
\mathbb{E}\left[|\widetilde\bg[t]^*\bh[t]|^2\right]  & = \mathbb{E}\left[\etaf^2\bh[t-1]^*\left[\widehat{\bg}[t-1],\ \ep \right]\bbeta\bbeta^*\left[\widehat{\bg}[t-1],\ \ep \right]^*\bh[t-1]+ \right. \\ 
& \qquad\qquad\qquad\qquad\qquad \left. 2\mathrm{Re}\left\{ \etaf\bar{\eta}_\mathrm{f}\bh[t-1]^*\left[\widehat{\bg}[t-1],\ \ep \right]\bbeta\bbeta^*\bx[t]\right\} +\bar{\eta}_\mathrm{f}^2\bx[t]^*\bbeta\bbeta^*\bx[t]\right] \nonumber \\
&\stackrel{(a)}{=}  \mathbb{E}\left[\etaf^2\bh[t-1]^*\left[\widehat{\bg}[t-1],\ \ep \right]\bbeta\bbeta^*\left[\widehat{\bg}[t-1],\ \ep \right]^*\bh[t-1]\right]+ \bar{\eta}_\mathrm{f}^2\frac{1}{L}\\
& \stackrel{(b)}{=} \etaf^2 \cos^2\left(\|\ep \|\right)\mathbb{E}\left[\left|\bh[t-1]^*\widehat{\bg}[t-1]\right|^2\right] + \etaf^2\sin^2\left(\|\ep \|\right)\mathbb{E}\left[\left|\bh[t-1]^*\ep \right|^2\right] \\ 
&\qquad\quad  + \mathbb{E}\left[2\mathrm{Re}\left\{\etaf^2\cos\left(\|\ep \|\right)\sin\left(\|\ep \|\right)\bh[t-1]^*\widehat{\bg}[t-1]\ep ^*\bh[t-1]\right\}\right] + \bar{\eta}_\mathrm{f}^2\frac{1}{L}, \nonumber
\end{split}
\end{align}
where $(a)$ follows because $\bx[t]$ is a zero-mean vector and independent of $\bh[t-1]^*\left[\widehat{\bg}[t-1],\ \ep \right]\bbeta\bbeta^*$ and $\mathbb{E}\left[\bx[t]^*\bbeta\bbeta^*\bx[t]\right]=\frac{1}{L}$. The expectation of the third term in $(b)$ is zero as a result of Assumption \ref{as:quant_err} and the fact that $\ep $ and $\widehat{\bg}[t-1]$ are instantaneously orthogonal. Moreover, leveraging the independence of $\|\bh[t-1]\|$ and $\bh[t-1]/\|\bh[t-1]\|$, and that $\mathbb{E}\left[\|\bh[t-1]\|^2\right]=1$, the objective function simplifies to
\begin{align}
\begin{split}
\mathbb{E}\left[|\widetilde\bg[t]^*\bh[t]|^2\right]  = \etaf^2 \cos^2\left(\|\ep \|\right)\mathbb{E}\left[\left|\bg[t-1]^*\widehat{\bg}[t-1]\right|^2\right] +\etaf^2\sin^2\left(\|\ep \|\right)\mathbb{E}\left[\left|\bg[t-1]^*\ep \right|^2\right] + \bar{\eta}_\mathrm{f}^2\frac{1}{L}. \nonumber
\end{split}
\end{align}
Letting $\widehat{\rho}=\bg[t-1]^*\widehat{\bg}[t-1]$, the channel $\bg[t-1]$ can be written as $\widehat{\rho}\widehat{\bg}[t-1]+\sqrt{1-|\widehat{\rho}|^2} \widetilde{\bee}[t-1]$ where $\widetilde{\bee}[t-1]$ is a unit norm tangent vector representing the quantization error at time $t-1$. By Assumption \ref{as:quant_err}, $\widetilde{\bee}[t-1]$ is uniformly distributed in the tangent space and independent of $|\widehat{\rho}|$, which results in
\begin{align}
\mathbb{E}\left[|\widetilde\bg[t]^*\bh[t]|^2\right]  & = \etaf^2 \cos^2\left(\|\ep \|\right)\mathbb{E}\left[|\widehat{\rho}|^2\right] +\etaf^2\sin^2\left(\|\ep \|\right)(1-\mathbb{E}\left[|\widehat{\rho}|^2\right])\mathbb{E}\left[\left|\widetilde{\bee}[t-1]^*\ep \right|^2\right] + \bar{\eta}_\mathrm{f}^2\frac{1}{L} \nonumber \\
& \stackrel{(a)}{=} \etaf^2 \cos^2\left(\|\ep \|\right)\mathbb{E}\left[|\widehat{\rho}|^2\right] +\etaf^2\sin^2\left(\|\ep \|\right)(1-\mathbb{E}\left[|\widehat{\rho}|^2\right])\frac{1}{L-1} + \bar{\eta}_\mathrm{f}^2\frac{1}{L}, \label{eqn:quant_error_18}
\end{align}
where $(a)$ is since the isotropic distribution of $\widetilde{\bee}[t-1]$ implies that the term $\left|\widetilde{\bee}[t-1]^*\ep \right|^2$ is beta distributed with mean $\mathbb{E}\left[\left|\widetilde{\bee}^*\ep \right|^2\right]=1/(L-1)$. Therefore the predictor's objective function is independent of the predicted direction of $\ep $. Further, $\mathbb{E}\left[|\widehat{\rho}|^2\right]$ is practically such that $\mathbb{E}\left[|\widehat{\rho}|^2\right]> (1-\mathbb{E}\left[|\widehat{\rho}|^2\right])$, i.e. the quantizer's output is expected to be a more accurate estimate of the channel than a random vector. Thus $\mathbb{E}\left[|\widehat{\rho}|^2\right]> (1-\mathbb{E}\left[|\widehat{\rho}|^2\right])/(L-1)$ and $\mathbb{E}\left[|\widetilde\bg[t]^*\bh[t]|^2\right]$ is maximized in (\ref{eqn:quant_error_18}) by letting $\|\ep \|=0$ which achieves
\begin{equation}
\mathbb{E}\left[|\widetilde\bg[t]^*\bh[t]|^2\right]  = \etaf^2\mathbb{E}\left[|\widehat{\rho}|^2\right]+\bar{\eta}_\mathrm{f}^2\frac{1}{L}.
\end{equation}
As a result, the predictor's objective function is maximized by selecting $\|\ep \|=0$ and thus $\widetilde{\bg}[t]=\widehat{\bg}[t-1]$. This coincides with the differential feedback strategy presented, thus the performance of the proposed framework can not be further improved by such a predictor for the channel model considered.

The derivation in this section, however, does not preclude potential benefits from more sophisticated prediction functions, especially when the band-limited nature of the channel's Doppler spectrum is exploited. While a rich body of literature on channel prediction does exist~\cite{komninakis2002multi, kho2008mimo,svantsson}, work on predicting processes that evolve on the Grassmann manifold is limited to~\cite{Takao2,jafarkhani} which use predictors and channel models similar to the ones considered in this section. The design of more sophisticated Grassmannian predictors that better exploit channel structure is left for future work.

\section{Design Considerations \& Codebook Construction}    \label{sec:gdc_design}

In this section, we discuss the proper initialization of the proposed algorithm and construct adaptive quantization codebooks for the tangent magnitude and direction. 

\subsection{Initialization}

The proper operation of the Grassmannian differential feedback algorithm is ensured by the fact that, at each iteration, both transmitter and receiver can calculate a common quantized channel based on the quantized tangent vector and the previous quantized channel. For the output of the algorithm to be the same at the transmitter and receiver, however, a common initial vector, $\widehat{\bg}[0]$, is required. Reinitialization can also be used to recover from feedback bit errors. This initial vector can, for example, be an estimate of $\widehat{\bg}[0]$ that is fed back from receiver to transmitter in the first iteration along with $\widehat{\bee}[0]$. Alternatively, it can be based on a memoryless quantization of the channel using methods such as \cite{love-heath-limited-feedback-unitary, Mukkavilli}. The additional overhead incurred is amortized when there are many subsequent feedback stages. Another option is to initialize the estimate $\widehat{\bg}[0]$ to a common random vector from which the algorithm will be able to recover. In this paper, we use memoryless quantization with a random vector codebook at $t=1$.

\subsection{Tangent Magnitude Quantization}

The tangent vector calculated in (\ref{eqn:tangent}) can be decomposed into a tangent magnitude and a unit norm tangent direction. 
We propose to quantize the tangent magnitude and direction separately, which ensures that search complexity remains manageable. The problem of quantizing the tangent magnitude is that of quantizing a non-negative scalar. We propose to quantize the tangent magnitude using a Euclidean distance distortion metric
\begin{equation}
\widehat{e}_\mathrm{mag}[t]= \argmin_{e_i \in \Cmag}\ \left|\|\bee[t]\|-e_{i}\right|,
\end{equation}
where $\Cmag $ is the magnitude quantization codebook. The index of the minimizer is then sent to the transmitter via a delay and error-free link that requires $\Nmag =\log(\left|\Cmag \right|)$ bits. A locally optimal quantization codebook can be found algorithmically using Lloyd's algorithm \cite{gray_vec_quant} given either the probability density function (pdf) of the tangent magnitude, or a training set of magnitudes which provide an empirical probability mass function. Finding the exact pdf of the tangent magnitude, however, has been intractable analytically thus far. Generating a good representative training set as input to the Lloyd algorithm is also difficult. The difficulty in both these approaches is that the magnitude pdf depends on the quantization process itself. That is to say that quantizing the tangent direction and magnitude itself could change their distribution in the next step. What is needed is therefore finding a pdf that is a fixed point of the quantization process, which has proven to be difficult. One solution which is adopted in~\cite{Takao2}  is to uniformly quantize a range of magnitudes, $\|\bee[t]\| \in \left[0, 1\right]$. Unfortunately, \cite{Takao2} has shown that quantization error in the magnitude creates an error floor beyond which CSI can not be improved in highly correlated channels where predictive feedback is most useful. In the limit of perfectly static channels, such magnitude quantization prevents feedback from converging to perfect CSI. 

We propose to adapt the quantization range, $[\alpha\emin [t],\ \beta\emax [t]]$, to the dynamics of the system and to uniformly quantize the magnitude in that range. To do so, the algorithm maintains a \textit{sliding window} of the quantization range of interest. This adaptive process maintains a running average of the tangent vector magnitude and uses it to update the lower and upper limits of the quantization range. The quantization range, or window, is calculated using quantities that are commonly available at both transmitter and receiver. As a result, both nodes can independently keep track of the current quantization range. 

The moving average of quantized tangent magnitudes is given by:
\begin{equation}
\eavg [t+1]  = \left(1-\frac{1}{\tau}\right)\eavg [t]+\tau\emag [t],
\end{equation}
where $\tau$ is the smoothing factor that can be adjusted by the system. 
At each iteration either limit of the quantization range is then updated as follows: 
\begin{eqnarray}
\emin [t+1] & = & \left\{ \begin{array}{cc} 
\left(1-\frac{1}{\tau}\right)\emin [t]+\frac{1}{\tau}\emag [t] \quad & \emag [t]\leq \eavg [t] \nonumber \\
\emin [t] \quad & \emag [t]\geq \eavg [t]
\end{array}\right.
\\
\emax [t+1] & = & \left\{\begin{array}{cc}

\emax [t] \quad & \emag [t] \leq \eavg [t] \nonumber \\
\left(1-\frac{1}{\tau}\right)\emax [t]+\frac{1}{\tau}\emag [t] \quad & \emag [t]\geq \eavg [t].
\end{array}\right.
\end{eqnarray}
This allows the feedback algorithm to accurately track the statistics of the tangent magnitude and quantize the current range of interest with higher resolution. In static channels, this allows the system to converge to perfect CSI removing the error floor caused by constant magnitude codebooks. 

The fixed scalars $\alpha$ and $\beta$ must now be such that $\alpha <1<\beta$ to enable the flexible adjustment of the window $[\emin [t],\ \emax [t]]$. To see this, note that to let $\emin [t]$ decrease, it is necessary to be able to quantize values less than $\emin [t]$. Similarly, allowing $\emax [t]$ to increase requires quantizing values greater than it. In simulation we use $\tau=5$, $\alpha=1/2$ and $\beta=2$. While the chosen parameter values yield good results as shown in Section \ref{sec:sims}, system performance can be further improved by optimizing over the values of $\tau$, $\alpha$, and $\beta$.


\subsection{Tangent Direction Quantization}

The problem of quantizing the tangent direction vector is that of quantizing a unit norm vector which lies in the tangent space orthogonal to the base vector $\widehat{\bg}[t-1]$, that is $\bee[t]^*\widehat{\bg}[t-1]=0$. Grassmannian codebooks often used for limited feedback \cite{love-heath-limited-feedback-unitary, Mukkavilli} are not suitable for tangent vector quantization since they do not enforce the structural constraint that requires the tangent direction codewords to be orthogonal to the base vector $\widehat{\bg}[t-1]$. If the quantized tangent vector is not orthogonal to the base vector $\widehat{\bg}[t-1]$, the geodesic path is undefined and the output of $G(\widehat{\bg}[t-1],\widehat{\bee}[t],\ell)$ does not lie on the manifold. Finally, note that the tangent space changes for each base vector $\widehat{\bg}[t-1]$ and thus the codebook must be adapted \cite{heath_prog_ref}, i.e., fixed codebooks should not be used. 

To respect the varying tangent space geometry and orthogonality constraints, we propose a canonical generating codebook that is adapted at each iteration; only the canonical codebook is actually stored. This canonical codebook quantizes the tangent space at a special $L \times 1$ base vector. The canonical codebook is rotated at each channel realization to match the geometry of the current tangent space.

\begin{definition}
The $\Ndir $-bit canonical tangent codebook is a codebook of $2^{\Ndir }$ unit norm vectors orthogonal to the special base vector $x_b \defeq \left[1, 0, \hdots, 0\right]$. Because of the structure of $x_b$, the generator codebook, $\Cdir(\bx_b)$, has elements, $\bv_\ell \in \Cdir(\bx_b)$, of the form $\bv_\ell= \left[0, \  \tilde{\bv}_\ell^*\right]^* $ where $\tilde{\bv}_\ell$ is an $(L-1) \times 1$ unit norm vector.
\end{definition}

The canonical codebook can be constructed by appending a leading $0$ to a codebook of $(L-1) \times 1$ vectors. The canonical codebook, which can be used to quantize the tangent space at $\bx_b$, can now be rotated at each iteration (or channel realization) to quantize the tangent space at a new vector $\bx$. We define this rotation as follows~\cite{jafarkhani,heath_prog_ref}.
\begin{definition}
Let $\bU(\bx): \mathbb{C}^{L \times 1} \mapsto \mathcal{U}^{L \times L}$ be the function that determines the unitary rotation matrix that rotates $\bx_b$ to $\bx$, i.e. $\bU(\bx)\bx_b=\bx$. 
\end{definition}
Perhaps the easiest way to find such a rotation is to consider the Householder matrix. Let $\bH(\bx,\bx_b)=\bI_L - \bu\bu^*/\bu^*\bx_b$ where $\bu=\bx_b-\bx$ \cite{householder-transform}. Note that $\bH(\bx,\bx_b)$ is a unitary matrix and that if we let $\bU(\bx)=\bH(\bx,\bx_b)$ then $\bU(\bx)\bx_b=\bx$ as required. Using the rotation operation it is possible to rotate the canonical codebook to the direction of any particular base point and generate tangent direction codebooks for an arbitrary channel. For a base point, $\widehat{\bg}[t-1]$, the quantizer chooses $\edir$ from the rotated codebook 
\begin{equation}
\edir \in \Cdir(\widehat{\bg}[t-1]) =\left\{\bU(\widehat{\bg}[t-1])\bv_1, \bU(\widehat{\bg}[t-1])\bv_2, \hdots, \bU(\widehat{\bg}[t-1])\bv_{2^{\Ndir }}\right\}.
\nonumber
\end{equation}
It can be verified that all vectors $\bU(\widehat{\bg}[t-1])\bv_\ell$ lie in the tangent plane at $\widehat{\bg}[t-1]$ since $\widehat{\bg}[t-1]^*\bU(\widehat{\bg}[t-1])\bv_\ell=\bx_b^*\bv_\ell=0, \ \forall \ell$.

The proposed codebook design allows us to rotate a canonical codebook to perfectly match the tangent plane at each iteration. This ensures that the output of the Grassmannian differential feedback algorithm remains on the manifold. Moreover, if changes in the channel are isotropic in the $L$ dimensional space, then so are the normalized tangent vectors, implying good performance with canonical codebooks that uniformly quantize the tangent space at $\bx_b$. In Section \ref{sec:sims}, good performance is achieved even with random vector canonical codebooks. In general, for a given correlation matrix $\bR_{\bh_{k,\ell}}$, good canonical codebooks can be generated using Lloyd's algorithm, and different codebooks can be stored for environments with different correlation properties.

To formalize the tangent direction quantization, recall that the estimated channel in the next iteration is $G(\widehat{\bg}[t-1],\widehat{\bee}[t],1)$. Given that the loss in sum rate is related to the chordal distance between the channel and its estimate, the quantized tangent direction is computed as
\small
\begin{align}
\begin{split}
\edir[t] & = \argmin_{\bx_i \in \Cdir(\widehat{\bg}[t-1])}d(G(\widehat{\bg}[t-1],\emag [t]\bx_i,1), \bg[t])  = \argmax_{\bx_i \in \Cdir(\widehat{\bg}[t-1])} |G(\widehat{\bg}[t-1],\emag [t]\bx_i,1)^* \bg[t]|^2,
\label{eqn:dir_quantization}
\end{split}
\end{align}
\normalsize
where the tangent magnitude, $\emag [t]$, is given by the output of the magnitude quantization step. We note that although (\ref{eqn:dir_quantization}) intuitively includes a rotation of the full canonical codebook, computational complexity can be reduced by noticing that an equivalent rotation can be applied to the channels $\widehat{\bg}[t-1]$ and $\bg[t]$ instead, and the codebook need not be actually rotated. The details of this simplification are omitted but can be seen by expanding the objective function $d(G(\widehat{\bg}[t-1],\emag [t]\bx_i,1)$.

\section{Analytical Performance Characterization} \label{sec:approximation}
A complete performance analysis of differential or predictive quantization is difficult due to its recursive nature and, in our case, due to the CSI's Grassmannian structure~\cite{love-utilizing-temporal-correlation}. Relevant performance analysis when the quantized information evolves on the Grassmannian manifold is limited to incomplete performance characterizations\cite{jafarkhani,Takao2}. The analysis of the Householder scheme in \cite{jafarkhani} is limited to characterizing the performance of an open loop predictor, similar to Section \ref{sec:no_pred}. The resulting lower bound on performance in \cite{jafarkhani} neglects the quantization process and is not a function of the number of feedback bits. The performance analysis in \cite{Takao2} presents partial distortion bounds that are independent of the channel's temporal correlation and Doppler spread. 

In this section we derive an approximation for the distortion achieved by the proposed quantization strategy. Unlike the analysis in \cite{jafarkhani, Takao2}, our characterization accounts for both the number of feedback bits and the channel's temporal correlation, thus capturing most of the fundamental system parameters. Given the distortion expression, we leverage our analysis in \cite{Ayach2010a} and the discussion in Section \ref{sec:imperfect_csi} to characterize IA's mean loss in sum rate when the proposed feedback strategy is used.

 We derive an approximation for the quantizer accuracy defined as $\mathbb{E}\left[|\bg[t]^*\widehat{\bg}[t]|^2\right]$, under three mild assumptions. First, we specialize our analysis to the case of channels with relatively low Doppler spread where differential feedback techniques yield the most gain~\cite{jafarkhani,love-utilizing-temporal-correlation}. In the low Doppler setting, we can make use of a small angle approximation~\cite{arias1998geometry, Takao2}. Second, we neglect the error introduced by the tangent magnitude quantizer of Section \ref{sec:gdc_design}. Effectively we make the following assumption.
\begin{assumption}
We assume perfect magnitude quantization, i.e. $\emag [t]=\|\bee[t]\|, \forall t$.
\label{as:mag_quant}
\end{assumption}
This is a reasonable assumption for our analysis as the numerical results in Section \ref{sec:csi_qual} indicate that the magnitude quantizer adopted achieves very low quantization error with as little as one quantization bit and approaches perfect quantization at all Doppler. Finally, the analysis re-invokes Assumption \ref{as:ar1} for simplicity of exposition. 


We start by expanding $\mathcal{D}(\etaf,\Ndir )=\mathbb{E}\left[|\bg[t]^*\widehat{\bg}[t]|^2\right]$ as
\small
\begin{align}
\begin{split}
\mathcal{D}(\etaf,\Ndir ) & =\mathbb{E}\left[\left|\left(\cos(\|\bee[t]\|)\widehat{\bg}[t-1]+ \sin(\|\bee[t]\|)\frac{\bee[t]}{\|\bee[t]\|}\right)^*\left(\cos(\emag [t]) \widehat{\bg}[t-1] + \sin(\emag [t])\edir[t]\right)\right|^2\right] \\
& = \mathbb{E}\left[\left|\cos(\|\bee[t]\|)\cos(\emag [t]) +  \sin(\|\bee[t]\|)\sin(\emag [t])\frac{\bee[t]^*\edir[t]}{\|\bee[t]\|}\right|^2\right].  \nonumber
\end{split}
\end{align}
\normalsize
By invoking Assumption \ref{as:mag_quant}, $\mathcal{D}(\etaf,\Ndir )$ simplifies to
\small
\begin{align}
\begin{split}
\mathcal{D}(\etaf,\Ndir ) & = \mathbb{E}\left[\cos^4(\|\bee[t]\|) + 2\cos^2(\|\bee[t]\|)\sin^2(\|\bee[t]\|)\mathrm{Re}\left\{\frac{\bee[t]^*\edir[t]}{\|\bee[t]\|}\right\} + \sin^4(\|\bee[t]\|)\left|\frac{\bee[t]^*\edir[t]}{\|\bee[t]\|}\right|^2\right].
\label{eqn:approximation2}
\end{split}
\end{align}
\normalsize
Recall that at each new channel realization the quantized tangent direction $\edir$ is selected to minimize the chordal distance between $\bg[t]$ and $\widehat\bg[t]$, equivalently maximizing $|\bg[t]^*\widehat{\bg}[t]|^2$. By invoking Assumption \ref{as:mag_quant}, maximizing $|\bg[t]^*\widehat{\bg}[t]|^2$ becomes equivalent to maximizing the instantaneous value of the right hand side of (\ref{eqn:approximation2}). The distortion function in (\ref{eqn:approximation2}), however, is new; it is neither a Euclidean quantizer's distortion function nor is it a Grassmannian quantizer's phase-invariant distortion function. As a result, neither Euclidean nor Grassmannian quantization results can be used to find an expression for $\mathcal{D}(\etaf,\Ndir )$. 

By analyzing a similar alternative quantization strategy, however, it is possible to gain insight into the behavior of (\ref{eqn:approximation2}) and derive an approximation for $\mathcal{D}(\etaf,\Ndir )$.
Instead of directly quantizing the tangent direction to maximize the instantaneous value of (\ref{eqn:approximation2}), we propose to split $\edir$ into two components, $\edir=e^{-j\widehat{\theta}_e[t]}\widehat{\bee}_{\mathcal{G}}[t]$. The variables $\widehat{\theta}_e[t]$ and $\widehat{\bee}_{\mathcal{G}}[t]$ are then quantized separately using $N_\theta$ and $N_\mathcal{G}$ bits. To keep the amount of feedback constant, we require $N_\theta + N_\mathcal{G}=\Ndir $. Such separation, naturally incurs an increase in distortion when compared to our original ``joint'' quantization strategy. Therefore, after deriving a distortion expression for the separate quantization strategy, we pick $N_\theta$ and $N_\mathcal{G}$ to optimize its performance which leads to a more accurate characterization of the original ``joint'' quantization strategy.

Inserting the proposed tangent decomposition into (\ref{eqn:approximation2}) yields 
\begin{align}
\begin{split}
\mathcal{D}(\etaf,\Ndir )  = \mathbb{E}\left[ \vphantom{\frac{1}{1}} \cos^4(\|\bee[t]\|) + 2\cos^2(\|\bee[t]\|)\sin^2(\|\bee[t]\|) \right. & \mathrm{Re}\left\{\frac{e^{-j\widehat{\theta}_e[t]} \bee[t]^*\widehat{\bee}_\mathcal{G}[t]}{\|\bee[t]\|}\right\} +\\ 
& \left. \sin^4(\|\bee[t]\|)\left|\frac{e^{-j\widehat{\theta}_e[t]}\bee[t]^*\widehat{\bee}_\mathcal{G}[t]}{\|\bee[t]\|}\right|^2 \right]
\end{split}
\end{align}
\begin{align}
\begin{split}
\hspace{10pt} \approx \mathbb{E}\left[ \vphantom{\frac{1}{1}} \cos^4(\|\bee[t]\|) + 2\cos^2(\|\bee[t]\|)\sin^2(\|\bee[t]\|) \right.  \left|\frac{\bee[t]^*\widehat{\bee}_\mathcal{G}[t]}{\|\bee[t]\|}\right| \cos\left(\angle \bee[t]^*\widehat{\bee}_{\mathcal{G}}[t] -\widehat{\theta}_e[t]\right) & +\\ 
& \hspace{-100pt} \left. \sin^4(\|\bee[t]\|)\left|\frac{\bee[t]^*\widehat{\bee}_\mathcal{G}[t]}{\|\bee[t]\|}\right|^2\right] ,
\label{eqn:approximation3}
\end{split}
\end{align}
where $\angle \bee[t]^*\widehat{\bee}_\mathcal{G}[t]$ is the phase of  the inner product $\bee[t]^*\widehat{\bee}_\mathcal{G}[t]$ and as a result $\mathrm{Re}\left\{e^{-j\widehat{\theta}_e[t]} \bee[t]^*\widehat{\bee}_\mathcal{G}[t]\right\}= \mathrm{Re}\left\{e^{j(\angle \bee[t]^*\widehat{\bee}_\mathcal{G}[t]-\widehat{\theta}_e[t])} \left|\bee[t]^*\widehat{\bee}_\mathcal{G}[t]\right|\right\} = |\bee[t]^*\widehat{\bee}_\mathcal{G}[t]|\cos\left(\angle \bee[t]^*\widehat{\bee}_\mathcal{G}[t] -\widehat{\theta}_e[t]\right)$. Due to the decomposition of $\edir$ the distortion function decouples, i.e. the quantization of $\edir$ decouples into a scalar quantization step to minimize $|\angle \bee[t]^*\widehat{\bee}_\mathcal{G}[t] -\widehat{\theta}_e[t]|$ and a Grassmannian quantization step to maximize $|\bee[t]^*\widehat{\bee}_\mathcal{G}[t]|$. Thanks to the decomposition, we can now use known results about scalar quantizers and Grassmannian quantizers to analyze the performance of our decomposed strategy.

Recalling Assumption \ref{as:ar1} and specializing the analysis to channels with relatively low Doppler as stated earlier, we apply a small angle approximation which can be used to see that $\bee[t]\approx \sqrt{1-\etaf^2}\bz[t]$ \cite{Takao2,arias1998geometry}. 
Therefore by Assumption \ref{as:ar1} and the small angle approximation, one can see that the tangent direction $\bee[t]/\|\bee[t]\|$ is isotropic and independent of $\|\bee[t]\|$ and that $\sin^2(\|\bee[t]\|)=\|\bee[t]\|^2=\|\sqrt{1-\etaf^2}\bz[t]\|^2$. Using this fact, along with the approximation $\mathbb{E}\left[x^2\right]\approx \mathbb{E}\left[x\right]^2$, which is accurate when $x$ has relatively low variance, gives
\begin{align}
\begin{split}
\mathcal{D}(\etaf,\Ndir ) \approx \etaf^4+ 2\etaf^2(1-\etaf^2)\mathbb{E}\left[\cos\left(\angle \bee[t]^*\widehat{\bee}_\mathcal{G}[t] -\widehat{\theta}_e[t]\right)\right] & \mathbb{E} \left[\left|\frac{\bee[t]^*\widehat{\bee}_\mathcal{G}[t]}{\|\bee[t]\|}\right|\right]+ \\ & \hspace{-10pt} (1-\etaf^2)^2\mathbb{E}\left[\left|\frac{\bee[t]^*\widehat{\bee}_\mathcal{G}[t]}{\|\bee[t]\|}\right|^2\right].
\label{eqn:approx4}
\end{split}
\end{align}
Now, using an $N_\theta$-bit uniform codebook over $[0,2\pi]$ for $\widehat{\theta}_e[t]$ yields 
\begin{equation}
\mathbb{E}\left[\cos\left(\angle \bee[t]^*\widehat{\bee}_\mathcal{G}[t] -\widehat{\theta}_e[t]\right)\right]=\frac{\sin(\pi/2^{N_\theta})}{\pi/2^{N_\theta}},
\label{eqn:theta}
\end{equation}
since the $\bee[t]$ is isotropic which implies that $\angle \bee[t]^*\widehat{\bee}_\mathcal{G}[t]$ is uniform~\cite{love-heath-limited-feedback-unitary}. Further, using an $L-1$ dimensional Grassmannian codebook to construct the canonical codebook for $\widehat{\bee}_\mathcal{G}[t]$ yields 
\begin{equation}
\left|\frac{\bee[t]^*\widehat{\bee}_\mathcal{G}[t]}{\|\bee[t]\|}\right|^2<1-\frac{4}{2^{N_\mathcal{G}/(L-2)}}.
\label{eqn:grassbound}
\end{equation}
The proof of (\ref{eqn:grassbound}) follows from the derivations in \cite{Mukkavilli} and is thus omitted. Inserting (\ref{eqn:theta}) and (\ref{eqn:grassbound}) into (\ref{eqn:approx4}) yields the final result which we summarize in the following proposition.
\begin{proposition}
Under Assumptions \ref{as:ar1} and \ref{as:mag_quant}, the accuracy achieved by the proposed Grassmannian differential feedback strategy with a $2^{\Ndir }$ direction codebook is 
\begin{align}
\begin{split}
\mathcal{D}(\etaf,\Ndir ) \approx & \max_{N_\theta,\ N_\mathcal{G}}\etaf^4+ 2\etaf^2(1-\etaf^2)\frac{\sin(\pi/2^{N_\theta})}{\pi/2^{N_\theta}}\sqrt{1-\frac{4}{2^{N_\mathcal{G}/(L-2)}}}+(1-\etaf^2)^2\left(1-\frac{4}{2^{N_\mathcal{G}/(L-2)}}\right)\\
& s.t.\quad  N_\theta+N_\mathcal{G}=\Ndir . \nonumber
\end{split}
\end{align}
The values of $N_\theta$ and $N_\mathcal{G}$ can be easily determined by a search over $\Ndir $ possibilities. 
\label{prop:final_approx}
\end{proposition}

Finally, we note that the approximation in Proposition \ref{prop:final_approx} can now shed light on the performance of IA with the proposed feedback strategy. For example, it can be used in an identical manner as in \cite{Ayach2010a} to evaluate the mean loss in IA sum rate. Since deriving IA's mean loss sum rate parallels our derivation in \cite{Ayach2010a}, and since the main results of \cite{Ayach2010a, Thukral2009} have been summarized via (\ref{eqn:jensen}) and (\ref{eqn:interference_bound}), we present the final result on sum rate loss and refer the reader to \cite{Ayach2010a} for more detail. When the differential feedback strategy, with $\Ndir $ bits for the tangent direction, is used to fulfill IA's CSI requirement in a channel with $\etaf= J_0(2\pi \fd  \Ts)$, the mean loss in sum rate is such that
\begin{equation}
\Delta \Rsum\apprle \sum\limits_{k}\frac{d_k}{\Nsc}\log_2\left(1+\frac{\Nsc P}{\sigman}\sum_{\ell} \frac{\rho_{k,\ell}\left(d_\ell-\delta_{\ell,k}\right)}{d_\ell}\left(1-\mathcal{D} (\etaf,\Ndir )\right)\right), \label{eqn:rate_loss}
\end{equation}
where $\delta_{\ell,k}$ is the Kronecker delta function. As a result of (\ref{eqn:rate_loss}), IA sum rate with differential feedback is such that $\Rsum\apprge R^\mathrm{ideal}_\mathrm{sum}-\Delta \Rsum$, where $R^\mathrm{ideal}_\mathrm{sum}$ is the sum rate achieved with perfect CSI. We note, however, that the characterization given in (\ref{eqn:rate_loss}) and derived using the methods \cite{Ayach2010a} is known to be rather loose especially when CSI quality is poor, mainly due to the use of Jensen's inequality. Tighter, yet more involved, sum rate characterizations can be derived using alternative methods as in \cite{Ayach2012}.
\section{Simulation Results} \label{sec:sims}

In this section we present numerical performance results. First, we characterize the performance of Grassmannian differential feedback in relation to Doppler and codebook size, and explore the trade-off between codebook size and refresh rate. Second, we demonstrate the performance of IA when CSI is obtained via the Grassmannian differential feedback strategy.

\subsection{Quality of CSI}   \label{sec:csi_qual}

\begin{figure}
\centering
\includegraphics[width=3.5in]{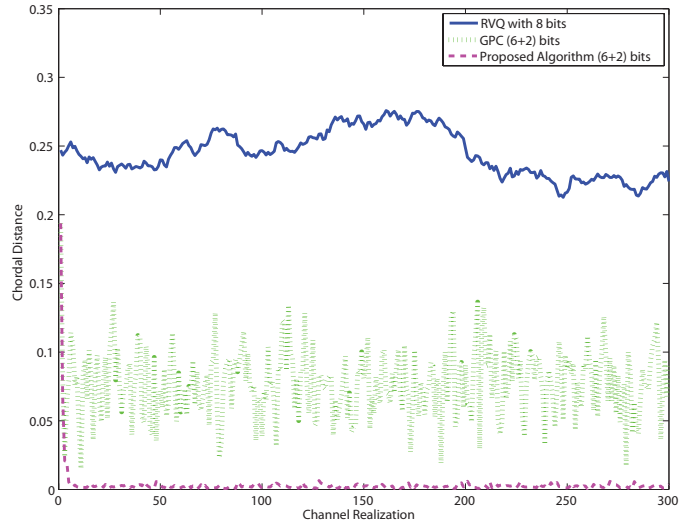}
\caption{Chordal distance, $d(\widehat{\bg}[t],\bg[t])$, plotted over time for a slowly channel with normalized Doppler $\fd \Ts=0.003$. This shows the high quality CSI achieved by the proposed algorithm with 6 and 2 bits for channel direction and magnitude respectively, compared to \cite{Takao2} and memoryless quantization with a random codebook of 8 bits.}
\label{fig:d_vs_time}
\end{figure}

First, we examine the quantization error resulting from using the proposed algorithm to quantize the channel's impulse response. We consider $L$ tap channels which are temporally correlated according to the $M$-order autoregressive channel model defined in~\cite{baddour2005autoregressive}. Unless otherwise stated, our simulations use $M=200$ and initialize the time series as in \cite{baddour2005autoregressive}. Although the analysis of Section \ref{sec:gdc} specialized $M=1$, adopting a more complex higher order model in simulation provides more realistic performance results. We restrict our attention to an uncorrelated uniform fading profile, that is $\mathbf{R_h}=\frac{1}{L}\bI_L$.The normalized channel vectors, $\frac{\bh_{k,\ell}[t]}{\|\bh_{k,\ell}[t]\|}$ constitute the correlated Grassmannian time series to be quantized.

\begin{figure}
\centering
\includegraphics[width=3.5in]{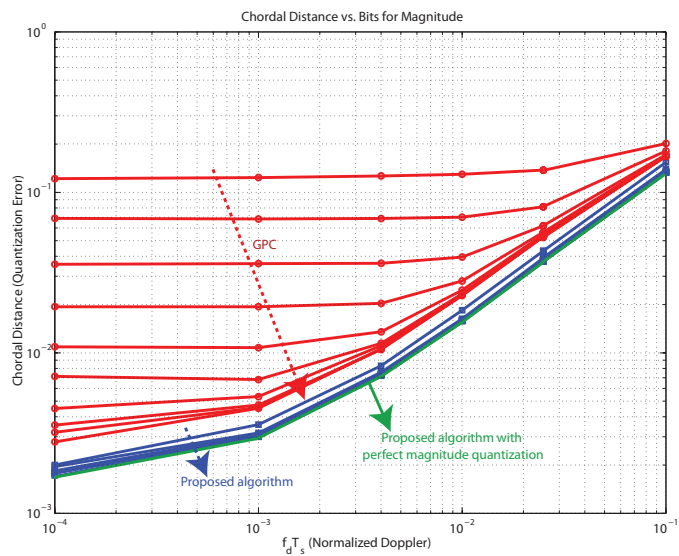}
\caption{This shows the average chordal distance, or quantization error, versus the number of bits allocated for the tangent magnitude (1 to 10 bits), while keeping direction bits fixed. This shows the good performance of the proposed algorithm when compared with \cite{Takao2}.}
\label{fig:d_vs_mbits}
\end{figure}

\begin{figure}
\centering
\includegraphics[width=3.5in]{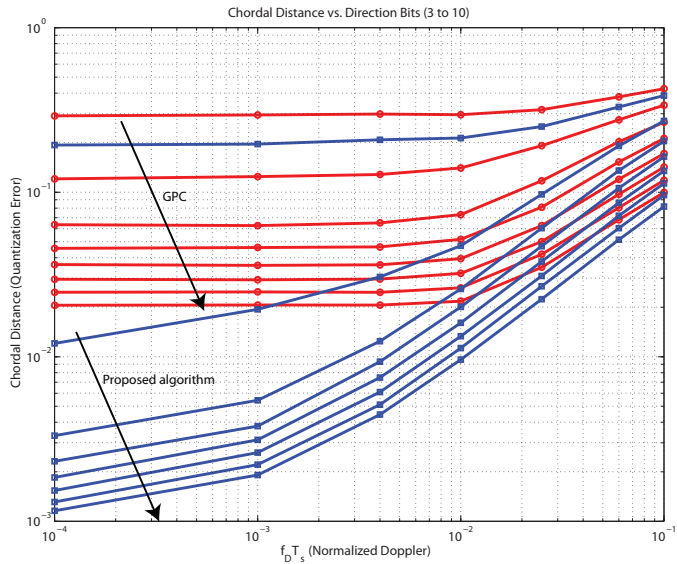}
\caption{This shows the average chordal distance, or quantization error, versus the number of bits allocated for the tangent direction, while keeping magnitude  bits fixed. This shows the good performance of the proposed algorithm when compared with \cite{Takao2}.}
\label{fig:d_vs_dirbits}
\end{figure}

Fig. \ref{fig:d_vs_time} shows the chordal distance between the actual and quantized channel for a three tap correlated channel with $\fd \Ts=0.005$. Comparing Grassmannian differential feedback to memoryless random vector quantization, we see that the proposed algorithm achieves a much lower chordal distance as early as the third or fourth iteration. The initial transient phase in which the quantization error is higher than the steady state is due to initializing the algorithm with an all ones vector, i.e. $\widehat{\bg}[0]=\frac{1}{\sqrt{3}}\left[1, 1, 1\right]^*$. The quick convergence to the steady state performance indicates that there is limited loss due to random initialization. The proposed algorithm also achieves a much lower steady state error compared to the GPC algorithm presented in \cite{Takao2} due to the adaptive codebooks presented. Examining the average quantization error achieved by both the proposed algorithm and the GPC algorithm of \cite{Takao2}, we see that the proposed algorithm consistently outperforms. This can be seen in Figs. \ref{fig:d_vs_mbits} and \ref{fig:d_vs_dirbits}.

Fig. \ref{fig:d_vs_mbits} plots the average error of both algorithms vs. Doppler spread for a varying number of magnitude quantization bits, $\Nmag $. The adaptive magnitude codebook, with even 1 bit of feedback, achieves 1.5dB lower distortion than the GPC algorithm with 10 bits of magnitude feedback. Adapting the codebook as presented in Section \ref{sec:gdc_design} allows accurate quantization and leaves very little to be gained from increasing the number of magnitude bits. Therefore, while \cite{Takao2} requires an increasing number of magnitude bits at low Doppler, which is the case of interest, our proposed algorithm works well with 1 bit for all Doppler spread. Moreover, if the total number of feedback bits is fixed, and the bits are allocated optimally across the direction and magnitude codebooks, it can be numerically shown that the optimal number of bits allocated to the magnitude is always 1. 

Fig. \ref{fig:d_vs_dirbits} plots the average error for a varying number of direction quantization bits, $\Ndir $, while keeping $\Nmag $ constant.  Fig. \ref{fig:d_vs_dirbits} again shows that the proposed algorithm consistently outperforms existing feedback schemes. We can see that the proposed algorithm can achieve up to 13dB lower quantization error than previous schemes. This is due to the proposed magnitude and tangent codebooks which adapt to the geometry of each point on the manifold. Moreover, Fig. \ref{fig:d_vs_dirbits} shows that, for the proposed algorithm, increasing the codebook size from 3 to 4, or 4 to 5 bits offers the greatest improvements in quantization error, after which only marginal improvements are experienced. Interestingly, as opposed to further increases in codebook size, going from 3 to 5 bits does not introduce a constant improvement at all Doppler and instead changes the slope of the curve providing much larger gains in slow fading channels.

\begin{table}[t!]
\centering
\caption{Minimum Direction Bits}
\begin{tabular}{|c|c|}
\hline Length of Channel & Suggested Direction Bits \\
\hline 2 & 3 \\
\hline 3 & 5 \\
\hline 4 & 7 \\
\hline 5 & 9 \\
\hline 6 & 11 \\
\hline
\end{tabular}
\label{table:bits}
\vspace{-0.15in}
\end{table}

Investigating this phase transition behavior further we notice that an error floor exists at low Doppler for all tangent direction codebooks smaller than 5 bits. Moreover, this phase transition is independent of the number of bits allocated to the magnitude, i.e. quantizing the magnitude with infinite precision does not change this behavior. Examining channels with more taps, we notice that this behavior is recurring, and that the phase transition occurs according to Table \ref{table:bits}. It is interesting to note that this phase transition behavior for an $L$-tap channel always occurs at $2L-1$ bits, which is the number of free variables in the tangent direction. A formal investigation of this is left for future work.

\begin{figure}
\centering
\includegraphics[width=3.5in]{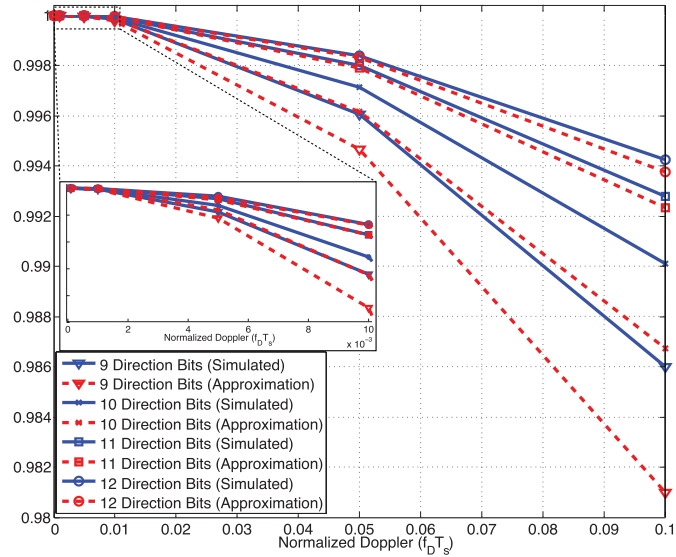}
\caption{This figure compares the simulated performance of differential feedback with the performance characterization derived in Section \ref{sec:approximation}. While the characterization is loose when feedback bits are low, due to the significant loss from separate quantization, the characterization becomes increasingly accurate as resolution increases.}
\label{fig:gdc_approx}
\end{figure}

Fig. \ref{fig:gdc_approx} plots $\mathbb{E}\left[|\bg[t]^*\widehat{\bg}[t]|^2\right]$ vs. $\fd \Ts$ for a practical range of $\Ndir $ to evaluate the accuracy of the performance characterization derived in Section \ref{sec:approximation}. Similarly to Section \ref{sec:approximation}, we consider the performance of the differential algorithm for a first order auto regressive channel model. As expected, the performance characterization is most accurate for slow fading channels and is reasonably accurate in faster fading scenarios. The reason for this is that the small angle approximation used is naturally suited for relatively low Doppler spreads. Moreover, the performance characterization gradually becomes more accurate as $\Ndir $ increases. Recall that the result was derived using a ``separate'' quantization scheme in which the tangent direction was decomposed into a phase angle, and a phase invariant direction. The bit allocation between the 2 variables, $N_\theta$ and $N_{\mathcal{G}}$, was then optimized to better characterize the superior ``joint quantization'' approach. The penalty due to such separate quantization is large when a small number of bits is available as the ``product'' of the two codebooks does not efficiently fill the tangent space. As the number of bits increases, the penalty due to such separate quantization diminishes and the performance characterization becomes more accurate even at high Doppler. Such behavior is in line with previous results on separate vs. joint quantization \cite{gray_vec_quant}.

\begin{figure}
\centering
\includegraphics[width=3.5in]{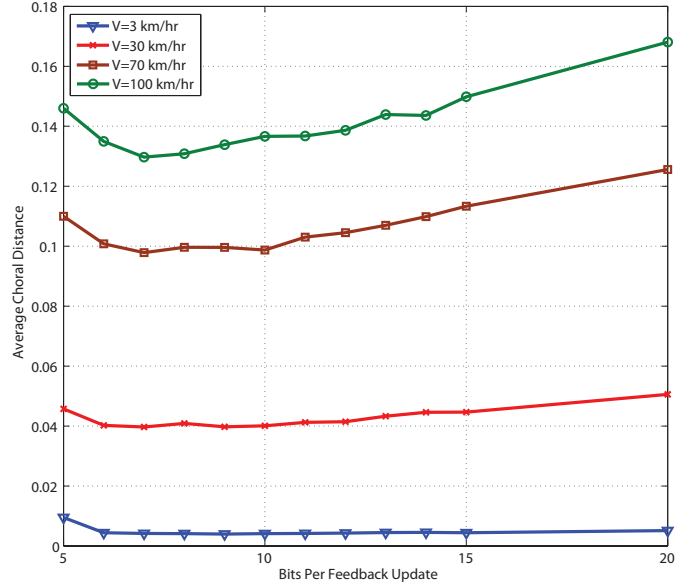}
\caption{Keeping the total feedback rate fixed at 5000 bits/s, this figure shows the average chordal distance or quantization error, versus the number of feedback bits. This shows that it is most often optimal to send frequent low resolution feedback. While slow fading channels can support and benefit from large feedback update periods, the performance in faster fading channels significantly deteriorates if refresh rate is sacrificed for codebook size. In all cases $\Nmag =1$ and $\Ndir $ is varied.}
\label{fig:refresh_rate}
\end{figure}

Finally, we highlight an issue that has often been overlooked in CSI feedback, with a few exceptions such as the analysis in \cite{love-refreshrate, huang2009limited} for single-user systems. Namely, the numerical analysis thus far has assumed a fixed normalized Doppler spread $\fd \Ts$, and a fixed number of feedback bits. Both these quantities, however, can be controlled by the system while keeping the overall overhead of feedback, i.e. feedback bit rate, fixed. When such optimization is possible, feedback frequency and codebook size can be traded-off depending on channel conditions, resulting in lower distortion. Fig. \ref{fig:refresh_rate} shows that increasing codebook size while keeping overall overhead constant can significantly deteriorate performance in fast fading channels. On the one hand, slow fading channels can support a large delay between high resolution feedback updates, since the channel does not change appreciably. On the other hand, decreasing the feedback time $\Ts$ to counter the high Doppler frequency $\fd $ yields better performance even if each update carries fewer bits. A formal analysis of this effect in the context of Grassmannian differential feedback is left for future work.

\subsection{IA Performance Analysis}  \label{sec:ia_sims}

In this section we present simulation results to demonstrate the performance of IA when channel knowledge at the transmitter is obtained via the Grassmannian differential feedback strategy detailed in Section \ref{sec:gdc}. We assume that the channels are correlated according to the same autoregressive model used in Section \ref{sec:csi_qual}, and that all the channels have equal Doppler spreads. Moreover, we assume that $\rho_{k,\ell}=\rho,\ \forall k,\ \ell$. Thus we define the signal-to-noise ratio as $SNR=\frac{\rho P}{\sigman}$.
Since the frequency extended system can be viewed as a virtual MIMO system, the sum rate achieved is calculated as~\cite{lozano2002capacity, Blum2003},
\begin{equation}
\Rsum = \sum_{k=1}^{K}\frac{1}{\Nsc} \log_{2} \left| \mathbf{I} + \frac{\Nsc P}{d_k}\left(\sigma^{2}\mathbf{I} + \frac{\Nsc P}{d_k}\sum_{m \neq k}\bH_{k,m}\widehat{\bF}_m\widehat{\bF}_m^*\bH_{k,m}^*  \right)^{-1} \left(\mathbf{H}_{k,k}\widehat{\mathbf{F}}_{k}\widehat{\mathbf{F}}_{k}^*\mathbf{H}_{kk}^*\right)\right|, \label{eqn:sim_sumrate}
\end{equation}
and the precoders, $\widehat{\bF}_k = \left[\widehat{\bff}_k^1, \widehat{\bff}_k^2, \hdots, \widehat{\bff}_k^{d_k}\right]$, are calculated given ideal or estimated CSI. We note here that the sum rate calculated in (\ref{eqn:sim_sumrate}) assumes that receivers use the perfect CSI available to them for decoding. For the results in this section, we use a modified version of the interference alignment algorithm proposed in \cite{Peters2010}, with the number of data symbols given by the conditions in \cite{Cadambe2008}. Although a closed form solution for the IA precoders exists for the single antenna frequency extended interference channel in \cite{Cadambe2008}, it has been shown that the solution in \cite{Cadambe2008} yields low sum rates if not further improved as in~\cite{torlak}. 

\begin{figure}
\centering
\includegraphics[width=3.5in]{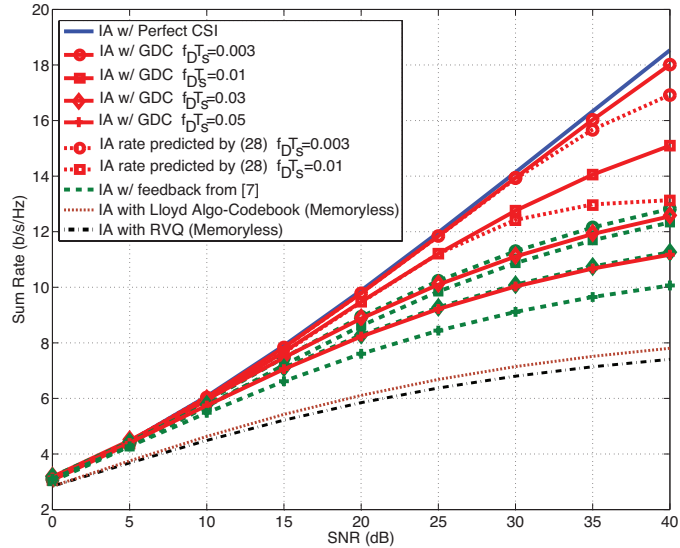}
\caption{This figure shows the performance of IA with imperfect CSI obtained through the proposed algorithm, the algorithm in \cite{Takao2}, as well as random vector quantization. This shows that for slowly varying channels, the proposed algorithm allows interference alignment networks to come very close to the perfect CSI upper bound.}
\label{fig:ia_with_gpc}
\end{figure}

Fig. \ref{fig:ia_with_gpc} shows the sum rate achieved by 3 users using IA over 16 channel extensions. CSI is obtained by our proposed algorithm, the strategy in \cite{Takao2}, and memoryless vector quantization. For all the feedback strategies shown, each 3-tap channel is quantized using 10 bits. For our feedback strategy and that of \cite{Takao2}, 7 and 3 bits are used for the tangent direction and magnitude respectively. From Fig. \ref{fig:ia_with_gpc}, one can show that, in the case of perfect CSI, the rate of increase of sum rate with SNR is 1.35, which approaches the 1.46 degrees of freedom predicted in theory over 16 subcarriers. 

As for the performance of IA with Grassmannian differential feedback, Fig. \ref{fig:ia_with_gpc} shows that the sum rate achieved by IA is significantly improved by exploiting temporal correlation in the channel. Although multiplexing gain will likely never be preserved with a constant number of feedback bits, we see that for correlated channels, the proposed algorithm exhibits close to perfect performance over a wide range of SNR. Practically, this is in fact what is needed. Since a system will almost never function in the asymptotically high SNR regime, the goal in practical systems is to optimize for the medium to high SNR regime where interference alignment is likely to be used. The proposed algorithm succeeds in providing negligible sum rate loss at up to 30dB, in channels with a normalized Doppler of up to $10^{-2}$. In fact this can be achieved with much less feedback bits as opposed to the 10 bit feedback performance shown in Fig. \ref{fig:ia_with_gpc}. The proposed algorithm continues to outperform memoryless quantization even at a significant Doppler of $\fd \Ts=0.05$ regardless of whether random codebooks (RVQ) or optimized codebooks, generated using the well-known Lloyd algorithm~\cite{gray_vec_quant, xia2005achieving}, are used. The proposed algorithm also consistently outperforms the GPC algorithm in \cite{Takao2} as a result of the adaptive magnitude and direction codebooks used. Finally, Fig. \ref{fig:ia_with_gpc} shows the analytical sum rate predicted by (\ref{eqn:rate_loss}) for the case of $\fd \Ts=0.003$ and $0.01$. As stated in Section \ref{sec:approximation}, the rate characterization is seen to be rather accurate for high SNRs up to 30dB, when 10 bits provide CSI of sufficient accuracy, but becomes progressively looser as SNR increases further and the CSI requirement becomes more stringent.


\section{Conclusion} \label{sec:conclusion}\
Limited feedback is a practical way to meet the CSI requirements of IA. To provide high resolution, we proposed a feedback strategy based on Grassmannian differential coding that uses the channel's temporal correlation to track channel responses by moving along geodesic paths defined via quantized tangent feedback. We showed that IA in conjunction with the proposed algorithm can benefit from more accurate channel knowledge and provide better sum rate performance for a large range of SNR conditions and a limited number of feedback bits. Moreover, our algorithm exhibits significant improvements over earlier methods of feeding back correlated time series on the manifold due to the optimized tangent and magnitude codebooks.

\vspace{-5pt}



\bibliographystyle{IEEEtran}
\bibliography{IEEEabrv,ia_gdc}

\begin{thebibliography}{10}
\providecommand{\url}[1]{#1}
\csname url@samestyle\endcsname
\providecommand{\newblock}{\relax}
\providecommand{\bibinfo}[2]{#2}
\providecommand{\BIBentrySTDinterwordspacing}{\spaceskip=0pt\relax}
\providecommand{\BIBentryALTinterwordstretchfactor}{4}
\providecommand{\BIBentryALTinterwordspacing}{\spaceskip=\fontdimen2\font plus
\BIBentryALTinterwordstretchfactor\fontdimen3\font minus
  \fontdimen4\font\relax}
\providecommand{\BIBforeignlanguage}[2]{{%
\expandafter\ifx\csname l@#1\endcsname\relax
\typeout{** WARNING: IEEEtran.bst: No hyphenation pattern has been}%
\typeout{** loaded for the language `#1'. Using the pattern for}%
\typeout{** the default language instead.}%
\else
\language=\csname l@#1\endcsname
\fi
#2}}
\providecommand{\BIBdecl}{\relax}
\BIBdecl

\bibitem{Cadambe2008}
V.~Cadambe and S.~Jafar, ``{Interference Alignment and Degrees of Freedom of
  the K-User Interference Channel},'' \emph{IEEE Transactions on Information
  Theory}, vol.~54, no.~8, pp. 3425--3441, Aug. 2008.

\bibitem{Peters2010}
S.~Peters and R.~W. Heath, Jr., ``Cooperative algorithms for {MIMO}
  interference channels,'' \emph{IEEE Transactions on Vehicular Technology},
  vol.~60, no.~1, pp. 206--218, Jan. 2011.

\bibitem{Gomadam2008}
K.~Gomadam, V.~Cadambe, and S.~Jafar, ``A distributed numerical approach to
  interference alignment and applications to wireless interference networks,''
  \emph{IEEE Transactions on Information Theory}, vol.~57, no.~6, pp.
  3309--3322, 2011.

\bibitem{MMSE-IA}
D.~Schmidt, C.~Shi, R.~Berry, M.~Honig, and W.~Utschick, ``Minimum mean squared
  error interference alignment,'' \emph{Conference Record of the 43rd Asilomar
  Conference on Signals, Systems and Computers}, pp. 1106--1110, Nov. 2009.

\bibitem{Berry-BidirectionalIA}
\BIBentryALTinterwordspacing
C.~{Shi}, R.~A. {Berry}, and M.~L. {Honig}, ``{Adaptive Beamforming in
  Interference Networks via Bi-Directional Training},'' \emph{ArXiv preprint
  arXiv:1003.4764}, Mar. 2010. [Online]. Available:
  \url{http://arxiv.org/abs/1003.4764}
\BIBentrySTDinterwordspacing

\bibitem{dimakis}
D.~Papailiopoulos and A.~Dimakis, ``Interference alignment as a rank
  constrained rank minimization,'' \emph{in Proc. of IEEE Global
  Telecommunications Conference}, pp. 1--6, Dec. 2010.

\bibitem{Choi2009}
S.~W. Choi, S.~Jafar, and S.-Y. Chung, ``On the beamforming design for
  efficient interference alignment,'' \emph{IEEE Communications Letters},
  vol.~13, no.~11, pp. 847--849, Nov. 2009.

\bibitem{santamaria-max-sum-rate}
I.~Santamar{\'\i}a, O.~Gonz{\'a}lez, R.~W. Heath, Jr., and S.~W. Peters,
  ``Maximum sum-rate interference alignment algorithms for {MIMO} channels,''
  \emph{in Proc. of IEEE Global Telecommunications Conference}, pp. 1--6, Dec.
  2010.

\bibitem{Tresch}
R.~Tresch, M.~Guillaud, and E.~Riegler, ``On the achievability of interference
  alignment in the {K}-user constant {MIMO} interference channel,'' \emph{in
  Proc. of IEEE/SP 15th Workshop on Statistical Signal Processing}, pp.
  277--280, Sep. 2009.

\bibitem{love2008overview}
D.~Love, R.~W. Heath, Jr., V.~Lau, D.~Gesbert, B.~Rao, and M.~Andrews, ``An
  overview of limited feedback in wireless communication systems,'' \emph{IEEE
  Journal on Selected Areas in Communications}, vol.~26, no.~8, pp. 1341--1365,
  Oct. 2008.

\bibitem{Peters2010a}
S.~Peters and R.~W. Heath, Jr., ``{Orthogonalization to reduce overhead in MIMO
  interference channels},'' \emph{in Proc. of International Zurich Seminar},
  pp. 126--129, March 2010.

\bibitem{guillaud2011interference}
M.~Guillaud and D.~Gesbert, ``Interference alignment in the partially connected
  {K}-user {MIMO} interference channel,'' \emph{Proc. of European Signal
  Processing Conference (EUSIPCO), Barcelona, Spain}, pp. 1--5, Sept. 2011.

\bibitem{Thukral2009}
H.~Bolcskei and I.~Thukral, ``Interference alignment with limited feedback,''
  \emph{in Proc. of IEEE International Symposium on Information Theory}, pp.
  1759--1763, Jul. 2009.

\bibitem{Krishnamachari2009}
R.~Krishnamachari and M.~Varanasi, ``Interference alignment under limited
  feedback for {MIMO} interference channels,'' \emph{in Proc. of IEEE
  International Symposium on Information Theory}, pp. 619--623, Jun. 2010.

\bibitem{Ayach2010a}
{O. El Ayach} and R.~W. Heath, Jr., ``Interference alignment with analog
  channel state feedback,'' \emph{IEEE Transactions on Wireless Communication},
  vol.~11, no.~2, pp. 626--636, Feb. 2012.

\bibitem{caire710multiuser}
G.~Caire, N.~Jindal, M.~Kobayashi, and N.~Ravindran, ``Multiuser {MIMO}
  achievable rates with downlink training and channel state feedback,''
  \emph{IEEE Transactions on Information Theory}, vol.~56, no.~6, pp.
  2845--2866, Jun. 2010.

\bibitem{lee2009mimo}
J.~Lee, J.~Han, and J.~Zhang, ``{MIMO technologies in 3GPP LTE and
  LTE-advanced},'' \emph{EURASIP Journal on Wireless Communications and
  Networking}, pp. 1--10, March 2009.

\bibitem{trivellato2008transceiver}
M.~Trivellato, F.~Boccardi, and H.~Huang, ``On transceiver design and channel
  quantization for downlink multiuser mimo systems with limited feedback,''
  \emph{IEEE Journal on Selected Areas in Communications}, vol.~26, no.~8, pp.
  1494--1504, 2008.

\bibitem{sacristn2010differential}
D.~Sacristan-Murga and A.~Pascual-Iserte, ``{Differential feedback of MIMO
  channel gram matrices based on geodesic curves},'' \emph{IEEE Transactions on
  Wireless Communications}, vol.~9, no.~12, pp. 3714--3727, Dec. 2010.

\bibitem{sacristn2009differential}
D.~Sacristan-Murga, F.~Kaltenberger, A.~Pascual-Iserte, and A.~Perez-Neira,
  ``{Differential feedback in MIMO communications: Performance with delay and
  real channel measurements},'' \emph{in Proc. of Workshop on Smart Antennas},
  pp. 1--8, 2009.

\bibitem{love-utilizing-temporal-correlation}
K.~Kim, I.~H. Kim, and D.~Love, ``Utilizing temporal correlation in multiuser
  {MIMO} feedback,'' \emph{in Proc. of 42nd Asilomar Conference on Signals,
  Systems and Computers}, pp. 121--125, Oct. 2008.

\bibitem{RohRao_efficientfeedback}
J.~Roh and B.~Rao, ``Efficient feedback methods for {MIMO} channels based on
  parameterization,'' \emph{IEEE Transactions on Wireless Communications},
  vol.~6, no.~1, pp. 282--292, Jan. 2007.

\bibitem{jafarkhani}
L.~Liu and H.~Jafarkhani, ``Novel transmit beamforming schemes for
  time-selective fading multiantenna systems,'' \emph{IEEE Transactions on
  Signal Processing}, vol.~54, no.~12, pp. 4767--4781, Dec. 2006.

\bibitem{Takao2}
\BIBentryALTinterwordspacing
T.~Inoue and R.~W. Heath, Jr., ``Grassmannian predictive coding for limited
  feedback in multiple antenna wireless systems,'' \emph{submitted to IEEE
  Transactions on Signal Processing}, 2011. [Online]. Available:
  \url{http://arxiv.org/abs/1105.5782}
\BIBentrySTDinterwordspacing

\bibitem{love-refreshrate}
T.~Kim, D.~Love, and B.~Clerckx, ``Does frequent low resolution feedback
  outperform infrequent high resolution feedback for multiple antenna
  beamforming systems?'' \emph{IEEE Transactions on Signal Processing},
  vol.~59, no.~4, pp. 1654--1669, Apr. 2011.

\bibitem{clarke1968statistical}
R.~H. Clarke, ``{A statistical theory of mobile-radio reception},'' \emph{Bell
  System Technical Journal}, vol.~47, no.~6, pp. 957--1000, 1968.

\bibitem{gesbert2010multi}
D.~Gesbert, S.~Hanly, H.~Huang, S.~Shamai~Shitz, O.~Simeone, and W.~Yu,
  ``Multi-cell mimo cooperative networks: A new look at interference,''
  \emph{IEEE Journal on Selected Areas in Communications}, vol.~28, no.~9, pp.
  1380--1408, 2010.

\bibitem{martinian-waterfilling}
\BIBentryALTinterwordspacing
E.~Martinian, ``{Waterfilling gains O (1/SNR) at high SNR}.'' [Online].
  Available: \url{http://www.csua.berkeley. edu/\~{} emin/research/wfill.pdf.}
\BIBentrySTDinterwordspacing

\bibitem{makouei-simple-sinr-characterization}
B.~Nosrat-Makouei, J.~Andrews, and R.~W. Heath, Jr., ``A simple {SINR}
  characterization for linear interference alignment over uncertain {MIMO}
  channels,'' \emph{in Proc. of IEEE Int. Symposium on Inf. Theory}, pp.
  2288--2292, Jun. 2010.

\bibitem{love-heath-limited-feedback-unitary}
D.~Love and R.~W. Heath, Jr., ``Limited feedback unitary precoding for spatial
  multiplexing systems,'' \emph{IEEE Transactions on Information Theory},
  vol.~51, no.~8, pp. 2967--2976, Aug. 2005.

\bibitem{Mukkavilli}
K.~Mukkavilli, A.~Sabharwal, E.~Erkip, and B.~Aazhang, ``On beamforming with
  finite rate feedback in multiple-antenna systems,'' \emph{IEEE Transactions
  on Information Theory}, vol.~49, no.~10, pp. 2562--2579, Oct. 2003.

\bibitem{limitedfeedforward}
C.-B. Chae, D.~Mazzarese, T.~Inoue, and R.~W. Heath, Jr., ``{Coordinated
  Beamforming for the Multiuser MIMO Broadcast Channel With Limited
  Feedforward},'' \emph{IEEE Transactions on Signal Processing}, vol.~56,
  no.~12, pp. 6044--6056, Dec. 2008.

\bibitem{packings-for-MIMO-BC}
A.~Ashikhmin and R.~Gopalan, ``Grassmannian packings for efficient quantization
  in {MIMO} broadcast systems,'' \emph{in Proc. of IEEE International Symposium
  on Information Theory}, pp. 1811 --1815, Jun. 2007.

\bibitem{arias1998geometry}
T.~Arias, A.~Edelman, and S.~Smith, ``{The geometry of algorithms with
  orthogonality constraints},'' \emph{SIAM J. Matrix Anal. Appl}, vol.~20, pp.
  303--353, 1998.

\bibitem{bhagavatula2010limited}
R.~Bhagavatula, R.~W. Heath, Jr., and B.~Rao, ``Limited feedback with joint
  {CSI} quantization for multicell cooperative generalized eigenvector
  beamforming,'' \emph{in Proc. of IEEE International Conference on Acoustics
  Speech and Signal Processing}, pp. 2838--2841, Mar. 2010.

\bibitem{baddour2005autoregressive}
K.~Baddour and N.~Beaulieu, ``{Autoregressive modeling for fading channel
  simulation},'' \emph{IEEE Transactions on Wireless Communications}, vol.~4,
  no.~4, pp. 1650--1662, Jul. 2005.

\bibitem{tulino2004random}
A.~Tulino and S.~Verd{\'u}, \emph{{Random matrix theory and wireless
  communications}}.\hskip 1em plus 0.5em minus 0.4em\relax Now Publishers Inc,
  2004.

\bibitem{komninakis2002multi}
C.~Komninakis, C.~Fragouli, A.~Sayed, and R.~Wesel, ``Multi-input multi-output
  fading channel tracking and equalization using {Kalman} estimation,''
  \emph{IEEE Transactions on Signal Processing}, vol.~50, no.~5, pp.
  1065--1076, May 2002.

\bibitem{kho2008mimo}
Y.~Kho and D.~Taylor, ``{MIMO} channel estimation and tracking based on
  polynomial prediction with application to equalization,'' \emph{IEEE
  Transactions on Vehicular Technology}, vol.~57, no.~3, pp. 1585--1595, May
  2008.

\bibitem{svantsson}
T.~Svantesson and A.~Swindlehurst, ``A performance bound for prediction of
  {MIMO} channels,'' \emph{IEEE Transactions on Signal Processing}, vol.~54,
  no.~2, pp. 520--529, Feb. 2006.

\bibitem{gray_vec_quant}
R.~Gray, ``Vector quantization,'' \emph{IEEE ASSP Magazine}, vol.~1, no.~2, pp.
  4--29, April 1984.

\bibitem{heath_prog_ref}
R.~W. Heath, Jr., T.~Wu, and A.~Soong, ``Progressive refinement for high
  resolution limited feedback multiuser {MIMO} beamforming,'' \emph{in Proc. of
  42nd Asilomar Conference on Signals, Systems and Computers}, pp. 743--747,
  Oct. 2008.

\bibitem{householder-transform}
K.-L. Chung and W.-M. Yan, ``The complex {Householder} transform,'' \emph{IEEE
  Transactions on Signal Processing}, vol.~45, no.~9, pp. 2374--2376, Sep.
  1997.

\bibitem{Ayach2012}
\BIBentryALTinterwordspacing
{O. El Ayach}, {A. Lozano}, and R.~W. Heath, Jr., ``On the overhead of
  interference alignment: Training, feedback, and cooperation,''
  \emph{submitted to IEEE Transactions on Wireless Communication}, Apr. 2012.
  [Online]. Available: \url{http://arxiv.org/abs/1204.6100}
\BIBentrySTDinterwordspacing

\bibitem{huang2009limited}
K.~Huang, R.~Heath, and J.~Andrews, ``Limited feedback beamforming over
  temporally-correlated channels,'' \emph{Signal Processing, IEEE Transactions
  on}, vol.~57, no.~5, pp. 1959--1975, 2009.

\bibitem{lozano2002capacity}
A.~Lozano and A.~Tulino, ``Capacity of multiple-transmit multiple-receive
  antenna architectures,'' \emph{IEEE Transactions on Information Theory},
  vol.~48, no.~12, pp. 3117--3128, 2002.

\bibitem{Blum2003}
R.~Blum, ``{MIMO} capacity with interference,'' \emph{IEEE Journal on Selected
  Areas in Communications}, vol.~21, no.~5, pp. 793--801, Jun. 2003.

\bibitem{torlak}
D.~Kim and M.~Torlak, ``Optimization of interference alignment beamforming
  vectors,'' \emph{IEEE Journal on Selected Areas in Communications}, vol.~28,
  no.~9, pp. 1425--1434, Dec. 2010.

\bibitem{xia2005achieving}
P.~Xia, S.~Zhou, and G.~Giannakis, ``Achieving the welch bound with difference
  sets,'' \emph{IEEE Transactions on Information Theory}, vol.~51, no.~5, pp.
  1900--1907, 2005.

\end{thebibliography}

\end{document}